\documentclass[11pt,final,onecolumn]{IEEEtran}

\pdfoutput=1 

\usepackage{amsmath,amssymb,amsfonts,mathtools,pdfsync}
\usepackage[]{graphicx}
\usepackage{sidecap}
\usepackage{amsfonts}
\usepackage{setspace, enumerate}
\usepackage{color}
\usepackage{amsmath,mathtools}
\usepackage{multirow,ifpdf,pdfpages}
\usepackage{algpseudocode,algorithm}
\usepackage{citesort,framed,footmisc}
\usepackage[]{todonotes}
\usepackage[margin=0pt,font=small]{caption}
\usepackage{cancel,subcaption,rviewport}

\newcommand{\mA}{\ensuremath{\mathbf A}}

\newcommand{\mW}{\ensuremath{\mathbf W}}

\newcommand{\va}{\ensuremath{\mathbf a}}

\newcommand{\vd}{\ensuremath{\mathbf d}}
\newcommand{\ve}{\ensuremath{\mathbf e}}

\newcommand{\vp}{\ensuremath{\mathbf p}}
\newcommand{\vq}{\ensuremath{\mathbf q}}

\newcommand{\vs}{\ensuremath{\mathbf s}}
\newcommand{\vu}{\ensuremath{\mathbf u}}

\newcommand{\vw}{\ensuremath{\mathbf w}}
\newcommand{\vx}{\ensuremath{\mathbf x}}
\newcommand{\vy}{\ensuremath{\mathbf y}}
\newcommand{\vz}{\ensuremath{\mathbf z}}

\DeclareMathOperator*{\argmax}{arg\,max}

\newcommand{\delx}{\ensuremath{\partial \mathbf{x}}}

\newcommand{\sign}[1]{\mathrm{sign}{\left(#1\right)}}									 
\begin{document}
\IEEEoverridecommandlockouts
\title{Fast and Accurate Algorithms for Re-Weighted $\ell_1$-Norm Minimization}

\author{M.~Salman~Asif~and~Justin~Romberg
\thanks{M. S. Asif and J. Romberg are with the School of Electrical and Computer Engineering, Georgia Institute of Technology, Atlanta, GA 30332, USA. Email: \{sasif,jrom\}@gatech.edu. This work was supported by ONR grant N00014-11-1-0459 and a grant from the Packard Foundation. }
\thanks{Manuscript submitted to the {\it IEEE Transactions on Signal Processing} on July 24, 2012. }
}

\maketitle

\begin{abstract}
To recover a sparse signal from an underdetermined system, we often solve a constrained $\ell_1$-norm minimization problem. In many cases, the signal sparsity and the recovery performance can be further improved by replacing the $\ell_1$ norm with a ``weighted'' $\ell_1$ norm.
Without any prior information about nonzero elements of the signal, the procedure for selecting weights is iterative in nature.  Common approaches update the weights at every iteration using the solution of a weighted $\ell_1$ problem from the previous iteration.

In this paper, we present two homotopy-based algorithms that efficiently solve reweighted $\ell_1$ problems.
First, we present an algorithm that quickly updates the solution of a weighted $\ell_1$ problem as the weights change. Since the solution changes only slightly with small changes in the weights, we develop a homotopy algorithm that replaces the old weights with the new ones in a small number of computationally inexpensive steps.
Second, we propose an algorithm that solves a weighted $\ell_1$ problem by adaptively selecting the weights while estimating the signal.
This algorithm integrates the reweighting into every step along the homotopy path by changing the weights according to the changes in the solution and its support,
allowing us to achieve a high quality signal reconstruction by solving a single homotopy problem.
We compare the performance of both algorithms, in terms of reconstruction accuracy and computational complexity, against state-of-the-art solvers and show that our methods have smaller computational cost. In addition, we will show that the adaptive selection of the weights inside the homotopy often yields reconstructions of higher quality.
\end{abstract}

\section{Introduction}
We consider the fundamental problem of recovering a signal from (possibly incomplete) linear, noisy measurements.
We observe $\vy \in \mathbb{R}^M$ as
\begin{equation}\label{eq:y=Ax+e}
\vy = \mA\bar\vx+\ve,
\end{equation}
where $\bar\vx \in \mathbb{R}^N$ is an unknown signal of interest that is measured through an $M\times N$ matrix $A$, and $\ve\in \mathbb{R}^M$ is noise. Recent work in compressive sensing and sparse approximation has shown that if $\bar\vx$ is sparse and $\mA$ obeys certain ``incoherence'' conditions, then a stable recovery is possible \cite{Candes_2006_CompressiveSampling,Donoho_2006_CS,Candes_2006_StableRecovery,CandesTao_2005_DecodingLP}. For instance, we can estimate $\bar \vx$ by solving the following convex optimization program:
\begin{equation}\label{eq:BPDN}
\underset{\vx}{\text{minimize}}\; \tau \|\vx\|_1 + \frac{1}{2}\|\mA\vx -\vy\|_2^2,
\end{equation}
where the $\ell_1$ term promotes sparsity in the solution, the $\ell_2$ term keeps the solution close to the measurements, and $\tau > 0$ is a user-selected regularization parameter.
The program \eqref{eq:BPDN}, commonly known as the LASSO or Basis Pursuit Denoising \cite{Tibshirani_1996_LASSO,Chen_99_BasisPursuit}, yields good numerical results in a variety of problems and comes with strong theoretical guarantees \cite{MeinshausenYu_2007_LassoType,CandesPlan_2008_NearIdealModelSelection,Chen_99_BasisPursuit,DonohoElad_2006_StableOvercomplete}.

Replacing the $\ell_1$ norm in \eqref{eq:BPDN} with a ``weighted'' $\ell_1$ norm can often enhance the sparsity of the solution and improve the signal recovery performance \cite{zou2006adaptive,candes_enhancing_2008,khajehnejad2010improved,chen2011penalized,CHA-2011-SPARS,FriedlanderMansour-2012-PartialSupp}. The weighted $\ell_1$-norm minimization form of \eqref{eq:BPDN} can be described as
\begin{equation}\label{eq:wtBPDN}
\underset{\vx}{\text{minimize}} \; \sum_{i=1}^N \vw_i |\vx_i| + \frac{1}{2}\|\mA\vx -\vy\|_2^2,
\end{equation}
where $\vw_i > 0$ denotes the weight at index $i$.
We can adjust the $\vw_i$ in \eqref{eq:wtBPDN} to selectively penalize different coefficients in the solution. To promote the same sparsity structure in the solution that is present in the original signal, we can select the $\vw_i$ such that they have small values on the nonzero locations of the signal and significantly larger values elsewhere.
Since information about the locations and amplitudes of the nonzero coefficients of the original signal is not available a priori, the critical task of selecting the weights is performed iteratively. Common approaches for such ``iterative reweighting'' re-compute weights at every iteration using the solution of \eqref{eq:wtBPDN} at the previous iteration.
Suppose $\widehat \vx$ denotes the solution of \eqref{eq:wtBPDN} for a given set of weights. For the next iteration, we compute the $\vw_i$ as
\begin{equation}\label{eq:wt_update}
\vw_i = \frac{\tau}{|\widehat \vx_i| + \epsilon},
\end{equation}
for $i=1,\ldots,N$, using an appropriate choice of positive values for parameters $\tau$ and $\epsilon$.
We use these updated weights in \eqref{eq:wtBPDN} to compute an updated signal estimate, which we then use in \eqref{eq:wt_update} to re-compute the weights for the next iteration.
%
The major computational cost of every iteration in such a reweighting scheme arises from solving \eqref{eq:wtBPDN}, for which a number of solvers are available \cite{BergFriedlander-2008-spgl1,Wright-2009-sparsa,beck-2009-FISTA,Becker_2011_NESTA,yang-2011-yall1,Becker_2011_TFOCS}. 

In this paper, we present two homotopy-based algorithms for efficiently solving reweighted $\ell_1$-norm minimization problems. In a typical homotopy algorithm for an $\ell_1$ problem, as the homotopy parameter changes, the solution moves along a piecewise-linear path, and each segment on this homotopy path is traced with a computationally inexpensive homotopy step.
%
The major computational cost for every homotopy step involves one full matrix-vector multiplication and one rank-one update of the inverse of a small matrix. A well-known example is the standard LASSO homotopy in which we trace a solution path for \eqref{eq:BPDN} by reducing the single parameter $\tau$ while updating the support of the solution by one element at every step \cite{OsbornePresnell_2000_NewApproachLasso,Efron_2004_LARS,AR_L1updating_JSTSP09}.
By comparison, \eqref{eq:wtBPDN} has $N$ parameters in the form of $\vw_i$, and both the homotopy algorithms we present in this paper change the $\vw_i$ in such a way that their respective solutions follow piecewise-linear paths in sequences of inexpensive homotopy steps.

First, we present an algorithm that quickly updates the solution of \eqref{eq:wtBPDN} as the weights change in the iterative reweighting framework.
Suppose we have the solution of \eqref{eq:wtBPDN} for a given set of weights $\vw_i$ and we wish to update the weights to $\tilde \vw_i$. We develop a homotopy program that updates the solution of \eqref{eq:wtBPDN} by replacing the old weights ($\vw_i$) with the new ones ($\tilde \vw_i$). Since the solution of \eqref{eq:wtBPDN} changes only slightly with small changes in the weights, the homotopy procedure utilizes information about an existing solution to update the solution in only a small number of inexpensive homotopy steps.

Second, we propose a new homotopy algorithm that performs an internal ``adaptive reweighting'' after every homotopy step. Our algorithm yields a solution for a weighted $\ell_1$ problem of the form \eqref{eq:wtBPDN} for which the final values of $\vw_i$ are not assigned a priori, but instead are adaptively selected inside the algorithm.
%
In our proposed homotopy algorithm, we follow a solution path for \eqref{eq:wtBPDN} by adaptively reducing each $\vw_i$, while updating the support of the solution by one element at every step.
After every homotopy step, we adjust the weights according to the changes in the support of the solution so that the $\vw_i$ on the support shrink at a faster rate, toward smaller values (e.g., of the form in \eqref{eq:wt_update}), while the $\vw_i$ elsewhere shrink at a slower rate, toward a predefined threshold ($\tau>0$).
This allows us to recover a high-quality signal by solving a single homotopy problem, instead of solving \eqref{eq:wtBPDN}
multiple times via iterative reweighting (i.e., updating $\vw_i$ after solving \eqref{eq:wtBPDN}). We have also observed that such an adaptive reweighting tends to provide better quality of reconstruction compared to the standard method of iterative reweighting.
In addition to assigning smaller weights to the active indices, this adaptive reweighting serves another purpose: it encourages active elements to remain nonzero, which in turn reduces the total number of homotopy steps required for solving the entire problem.

Our proposed adaptive reweighting method bears some resemblance to a variable selection method recently presented in \cite{Radchenko-2011-FLASH}, which adjusts the level of shrinkage at each step (that is equivalent to reducing the $\vw_i$ toward zero) so as to optimize the selection of the next variable. However, the procedure we adopt for the selection of $\vw_i$ in this paper is more flexible, and it offers an explicit control over the values of $\vw_i$, which we exploit to embed a reweighted $\ell_1$-norm regularization inside the homotopy.
%

The paper is organized as follows.
In Section~\ref{sec:algo}, we briefly discuss the homotopy algorithm for \eqref{eq:BPDN}, on which we then build our discussion of the two homotopy algorithms for solving reweighted $\ell_1$ problem.
In Section~\ref{sec:exp}, we present numerical experiments that compare performance of our proposed algorithms, in terms of reconstruction accuracy and computational complexity, against three state-of-the-art solvers for which we use old solutions as warm start during iterative reweighting.


\section{Algorithms}\label{sec:algo}
\subsection{LASSO homotopy}\label{sec:BPDN}
The well-known LASSO homotopy algorithm solves \eqref{eq:BPDN} for one desired value of $\tau$ by tracing the entire solution path for a range of decreasing values of $\tau$
(i.e., any point on the so-called homotopy path is a solution of \eqref{eq:BPDN} for a certain value of $\tau$)
\cite{OsbornePresnell_2000_NewApproachLasso,Efron_2004_LARS}.
Starting with a large value of $\tau$, LASSO homotopy shrinks $\tau$ toward its final value in a sequence of computationally inexpensive steps.
The fundamental insight is that as $\tau$ changes, the solution of \eqref{eq:BPDN} follows a piecewise-linear path in which the length and the direction of each segment is completely determined by the support and the sign sequence of the solution on that segment. This fact can be derived by analyzing the KKT optimality conditions for \eqref{eq:BPDN}, as given below in \eqref{eq:BPDN_opts}~\cite{Fuchs_2004_OnSparseRep,Donoho_2006_FastLl1}. The support of the solution changes only at certain critical values of $\tau$, when either a new nonzero element enters the support or an existing nonzero element shrinks to zero.
These critical values of $\tau$ are easy to calculate at any point along the homotopy path.
For every homotopy step, we jump from one critical value of $\tau$ to the next while updating the support of the solution, until $\tau$ has been lowered to its desired value.

In every homotopy step, the update direction and the step-size for moving to a smaller critical value of $\tau$ can be easily calculated using certain optimality conditions, which can be derived using the subgradient of the objective in \eqref{eq:BPDN} \cite{Boyd_book_ConvexOptimization,Fuchs_2004_OnSparseRep}. At any given value of $\tau$, the solution $\vx^*$ for \eqref{eq:BPDN} must satisfy the following optimality conditions:
\begin{subequations}\label{eq:BPDN_opts}
\begin{gather}
\mA^T_\Gamma(\mA\vx^*-\vy) = -\tau \vz \label{eq:BPDN_opts_a} \\
\|\mA^T_{\Gamma^c}(\mA\vx^*-\vy)\|_\infty < \tau, \label{eq:BPDN_opts_b}
\end{gather}
\end{subequations}
where $\Gamma$ denotes the support of $\vx^*$, $\vz$ denotes the sign sequence of $\vx^*$ on $\Gamma$, and $\mA_\Gamma$ denotes a matrix with columns of $\mA$ at indices in the set $\Gamma$.
The optimality conditions in \eqref{eq:BPDN_opts} can be viewed as $N$ constraints that the solution $\vx^*$ needs to satisfy (with equality on the support $\Gamma$ and strict inequality elsewhere).
As we reduce $\tau$ to $\tau-\delta$, for a small value of $\delta$, the solution moves in a direction $\delx$, which to maintain optimality must obey
\begin{subequations}\label{eq:BPDN_update}
\begin{gather}
\mA^T_\Gamma(\mA\vx^*-\vy) + \delta \mA^T_\Gamma\mA\delx = -(\tau-\delta) \vz \label{eq:BPDN_update1} \\
\|\underbrace{\mA^T(\mA\vx^*-\vy)}_{\vp} + \delta \underbrace{\mA^T\mA\delx}_{\vd}\|_\infty \le (\tau-\delta). \label{eq:BPDN_update2}
\end{gather}
\end{subequations}
The update direction that keeps the solution optimal as we change $\delta$ can be written as
\begin{equation}\label{eq:BPDN_delx}
\delx = \begin{cases} (\mA^T_\Gamma \mA_\Gamma)^{-1} \vz & \text{on } \Gamma \\
0 & \text{otherwise}.
\end{cases}
\end{equation}
We can move in direction $\delx$ until one of the constraints in \eqref{eq:BPDN_update2} is violated, indicating we must add an element to the support $\Gamma$, or one of the nonzero elements in $\vx^*$ shrinks to zero, indicating we must remove an element from $\Gamma$.
The smallest step-size that causes one of these changes in the support can be easily computed as $\delta^* = \min(\delta^+,\delta^-)$, where
\begin{subequations}\label{eq:BPDN_delta}
\begin{flalign}
&& \delta^+ &= \min_{i\in \Gamma^c} \left(\frac{\tau-\vp_i}{1+\vd_i}, \frac{-\tau-\vp_i}{-1+\vd_i}\right)_+&&\\
\text{and}&& \delta^- &= \min_{i\in \Gamma} \left(\frac{-\vx^*_i}{\delx_i}\right)_+,&&
\end{flalign}
\end{subequations}
and $\min(\cdot)_+$ means that the minimum is taken over only positive arguments.
$\delta^+$ is the smallest step-size that causes an inactive constraint to become active at index $\gamma^+$, indicating that $\gamma^+$ should enter the support, and $\delta^-$ is the smallest step-size that shrinks an existing element at index $\gamma^-$ to zero.
The new critical value of $\tau$ becomes $\tau -\delta^*$ and the new signal estimate $\vx^*$ becomes $\vx^* + \delta^* \delx$, and its support and sign sequence are updated accordingly. We can now recompute the update direction and the step-size that define the next step in the homotopy path and the consequent one-element change in the support. We repeat this procedure until $\tau$ has been lowered to its desired value.

The main computational cost of every homotopy step comes from computing $\delx$ by solving an $S\times S$ system of equations in \eqref{eq:BPDN_delx} (where $S$ denotes the size of the support $\Gamma$) and from computing the vector $\vd$ in \eqref{eq:BPDN_opts_b} that is used to compute the step-size $\delta$ in \eqref{eq:BPDN_delta}. Since we know the values of $\vd$ on $\Gamma$ by construction and $\delx$ is nonzero only on $\Gamma$, the cost for computing $\vd$ is same as one application of an $M\times N$ matrix. Moreover, since $\Gamma$ changes by a single element at every homotopy step, instead of solving the linear system in \eqref{eq:BPDN_delx} from scratch, we can efficiently compute $\delx$ using a rank-one update at every step:
\renewcommand{\labelitemi}{{\tiny$\rhd$}}
\begin{itemize}
\item {\it Update matrix inverse:}
    We can derive a rank-one updating scheme by using matrix inversion lemma to explicitly update the inverse matrix $(\mA_\Gamma^T\mA_\Gamma)^{-1}$, which has an equivalent cost of performing one matrix-vector product with an
    $M\times S$ and one with an $S \times S$ matrix and adding a rank-one matrix to $(\mA^T_\Gamma\mA_\Gamma)^{-1}$. The update direction $\delx$ can be recursively computed with a vector addition. The total cost for rank-one update is approximately $MS+2S^2$ flops.
\item {\it Update matrix factorization:}
    Updating the inverse of matrix often suffers from numerical stability issues, especially when $S$ becomes closer to $M$ (i.e, the number of columns in $\mA_\Gamma$ becomes closer to the number of rows). In general, a more stable approach is to update a Cholesky factorization of $\mA_\Gamma^T\mA_\Gamma$ (or a QR factorization of $\mA_\Gamma$) as the support changes~\cite[Chapter~12]{Golub_1996_MatrixComputation}, \cite[Chapter~3]{Bjorck_1996_NumericalLS_book}. The computational cost for updating Cholesky factors and $\delx$ involves nearly $MS+3S^2$ flops.
\end{itemize}
As such, the computational cost of a homotopy step is close to the cost of one application of each $\mA$ and $\mA^T$ (that is, close to $MN+MS+3S^2+O(N)$ flops, assuming $S$ elements in the support).

\subsection{Iterative reweighting via homotopy}\label{sec:rwt}
In this section, we present a homotopy algorithm for iterative reweighting that quickly updates the solution of \eqref{eq:wtBPDN} as the weights $\vw_i$ change. Suppose we have solved \eqref{eq:wtBPDN} for a given set of $\vw_i$ and now we wish to solve the following problem:
\begin{equation}\label{eq:wtBPDN_tilde}
\underset{\vx}{\text{minimize}}\; \sum_i \tilde \vw_i|\vx_i|+ \frac{1}{2}\|\mA\vx-\vy\|_2^2,
\end{equation}
where the $\tilde \vw_i$ are the new weights.
%
%
To incorporate changes in the weights (i.e., replace the $\vw_i$ with the $\tilde \vw_i$) and quickly compute the new solution of \eqref{eq:wtBPDN}, we propose the following homotopy program:
\begin{equation}\label{eq:rwtBPDN}
\underset{\vx}{\text{minimize}}\; \sum_i ((1-\epsilon)\vw_i + \epsilon \tilde \vw_i)|\vx_i|+ \frac{1}{2}\|\mA\vx-\vy\|_2^2,
\end{equation}
where $\epsilon$ denotes the homotopy parameter that we change from zero to one to phase in the new weights and phase out the old ones. As we increase $\epsilon$, the solution of \eqref{eq:rwtBPDN} follows a homotopy path from the solution of \eqref{eq:wtBPDN} to that of \eqref{eq:wtBPDN_tilde}.
We show below that the path the solution takes is also piecewise linear with respect to $\epsilon$, making every homotopy step computationally inexpensive. The pseudocode outlining the important steps is presented in Algorithm~\ref{alg:rwt}.

\begin{algorithm}[t]
  \caption{Iterative reweighting via homotopy
    \label{alg:rwt}}
  \begin{algorithmic}[1]
    \Require{$\mA$, $\vy$, $\widehat \vx$, $\vw$, and $\tilde \vw$}
    \Ensure{$\vx^*$}
    \Statex
    \State {\bf Initialize: }$\epsilon = 0$, $\vx^* \gets \widehat\vx$
    \Repeat
    \State Compute $\delx$ in \eqref{eq:rwt_delx}   \Comment{Update direction}
    \State Compute $\vp, \vd, \vq$, and $\vs$ in \eqref{eq:rwt_optUpdate2}
    \State Compute $\delta^*$ in \eqref{eq:rwt_delta} \Comment{Step size}
    \State $\vx^* \gets \vx^* +\delta \delx$ \Comment{Update the solution}
    \If{$\delta^* = \delta^-$}
        \State $\Gamma \gets \Gamma \backslash \gamma^-$ \Comment{Remove an element from the support}
    \Else
        \State $\Gamma \gets \Gamma \cup \gamma^+$ \Comment{Add a new element to the support}
    \EndIf
    \Until{$\epsilon = 1$}
  \end{algorithmic}
\end{algorithm}

At any value of $\epsilon$, the solution $\vx^*$ must obey the following optimality conditions:
\begin{subequations}
\begin{flalign}
&&\va_i^T(\mA\vx^*-\vy) &= -((1-\epsilon)\vw_i + \epsilon \tilde \vw_i) \vz_i, ~ \text{for all } i \in \Gamma, && \\
\text{and}&&|\va_i^T(\mA\vx^*-\vy)| &< (1-\epsilon)\vw_i + \epsilon \tilde \vw_i, ~ \text{for all } i \in \Gamma^c,&&
\end{flalign}
\end{subequations}
where $\va_i$ denotes $i$th column of $\mA$.
As we increase $\epsilon$ to $\epsilon + \delta$, for some small $\delta$, the solution moves in a direction $\delx$  and the optimality conditions change as
\begin{subequations}\label{eq:rwt_optUpdate}
\begin{gather}
\mA_\Gamma^T(\mA\vx^*-\vy) + \delta \mA_\Gamma^T \mA\delx =   -((1-\epsilon)\mW + \epsilon \widetilde\mW)\vz  + \delta (\mW- \widetilde \mW)\vz \label{eq:rwt_optUpdate1}\\
|\underbrace{\va_i^T(\mA\vx^*-\vy)}_{\vp_i} + \delta  \underbrace{\va_i^T\mA\delx}_{\vd_i} | \le \underbrace{(1-\epsilon)\vw_i + \epsilon \tilde \vw_i}_{\vq_i} + \delta \underbrace{(\tilde \vw_i- \vw_i)}_{\vs_i}, \label{eq:rwt_optUpdate2}
\end{gather}
\end{subequations}
where $\mW$ and $\widetilde \mW$ denote $|\Gamma|\times |\Gamma|$ diagonal matrices with their diagonal entries being the values of $\vw$ and $\tilde \vw$ on $\Gamma$, respectively. The update direction is specified by the new optimality conditions \eqref{eq:rwt_optUpdate1} as
\begin{equation}\label{eq:rwt_delx}
\delx =
\begin{cases}
(\mA_\Gamma^T\mA_\Gamma)^{-1} (\mW-\widetilde \mW)\vz & \text{on } \Gamma \\
0 & \text{on } \Gamma^c.
\end{cases}
\end{equation}
As we increase $\delta$, the solution moves in the direction $\delx$ until either a new element enters the support of the solution (when a constraint in \eqref{eq:rwt_optUpdate2} becomes active) or an existing element shrinks to zero.
The stepsize that takes the solution to such a critical value of $\epsilon$ can be computed as $\delta^* = \min(\delta^+,\delta^-)$, where
\begin{subequations}\label{eq:rwt_delta}
\begin{gather}
\delta^+ = \min_{i\in \Gamma^c} \left(\frac{\vq_i- \vp_i}{-\vs_i+\vd_i}, \frac{-\vq_i - \vp_i}{\vs_i+\vd_i}\right)_+ \\
\delta^- = \min_{i\in \Gamma} \left(\frac{-\vx^*_i}{\delx_i}\right)_+.
\end{gather}
\end{subequations}
$\delta^+$ denotes the smallest step-size that causes a constraint in \eqref{eq:rwt_optUpdate2} to become active, indicating entry of a new element at index $\gamma^+$ in the support, whereas $\delta^-$ denotes the smallest step-size that shrinks an existing element at index $\gamma^-$ to zero.
The new critical value of $\epsilon$ becomes $\epsilon+\delta^*$, the signal estimate $\vx^*$ becomes $\vx^* + \delta^* \delx$, where its support and sign sequence are updated accordingly.
At every homotopy step, we jump from one critical value of $\epsilon$ to the next while updating the support of the solution, until $\epsilon = 1$.

The main computational cost of every homotopy step comes from solving a $|\Gamma|\times |\Gamma|$ system of equations to compute $\delx$ in \eqref{eq:rwt_delx} and one matrix-vector multiplication to compute the $\vd_i$ in \eqref{eq:rwt_delta}. Since $\Gamma$ changes by a single element at every homotopy step, the update direction can be computed using a rank-one update. As such, the computational cost of each homotopy step is close to one matrix-vector multiplication with $\mA$ and one with $\mA^T$.
We demonstrate with experiments in Sec.~\ref{sec:exp} that as the $\vw_i$ change, our proposed homotopy algorithm updates the solution in a small number of homotopy steps, and the total cost for updating weights is just a small fraction of the cost for solving \eqref{eq:wtBPDN} from scratch.

\subsection{Adaptive reweighting via homotopy}\label{sec:adp}
In this section, we present a homotopy algorithm that solves a weighted $\ell_1$-norm minimization problem of the form \eqref{eq:wtBPDN} by adaptively selecting the weights $\vw_i$. The motivation for this algorithm is to perform reweighting at every homotopy step by updating the weights according to the changes in the solution and its support.
Recall that in the standard LASSO homotopy we build the solution of \eqref{eq:BPDN} by adding or removing one element in the support while shrinking a single homotopy parameter $\tau$ and .
By comparison, each $\vw_i$ in \eqref{eq:wtBPDN} can act as a separate homotopy parameter, and we can attempt to achieve desired values for $\vw_i$ by adaptively shrinking them at every homotopy step.

In adaptive reweighting, we trace a solution path for \eqref{eq:wtBPDN} by adaptively reducing the $\vw_i$ while updating the solution and its support in a sequence of inexpensive homotopy steps.
At every homotopy step, we start with a solution of \eqref{eq:wtBPDN} for certain values of $\vw_i$.
We encourage the algorithm to focus on the set of active indices in the solution (i.e., the support of the solution) and reduce the $\vw_i$ so that they decrease at a faster rate and achieve smaller values on the active set than on the inactive set. Suppose, using certain criterion, we select the $\tilde \vw_i$ as the desired values of the weights.
As we change the $\vw_i$ toward the $\tilde \vw_i$, the solution moves in a certain direction until either the $\vw_i$ become equal to the $\tilde \vw_i$ or the support of the solution changes by one element. By taking into account any change in the support, we revise the values of $\tilde \vw_i$ for the next homotopy step. We repeat this procedure until each $\vw_i$ is reduced below a certain predefined threshold $\tau>0$.
%

In summary, we solve a single homotopy problem that builds the solution for a weighted $\ell_1$ problem of the form \eqref{eq:wtBPDN} by adjusting the $\vw_i$ according to the changes in the support of the solution.
A pseudocode with a high-level description is presented in Algorithm~\ref{alg:adp}.
Details regarding weights selection, update direction, and step size and support selection are as follows.

\begin{algorithm}[t]
  \caption{Adaptive reweighting via homotopy
    \label{alg:adp}}
  \begin{algorithmic}[1]
    \Require{$\mA$, $\vy$ and $\tau$}
    \Ensure{$\vx^*$, $\vw$}
    \Statex
    \State {\bf Initialize: }{$\vx^* \gets \mathbf{0},~\vw_i \gets \max_i|\va_i^T\vy|$ for all $i,~\Gamma \gets \argmax_i{|\va_i^T\vy|}$}
    \Repeat
    \State {Select $\tilde \vw_i$} \label{alg:adp_wtUpdate}
    \Comment{Desired values for the weights}
    \State Compute $\delx$ in \eqref{eq:adp_delx}   \Comment{Update direction}
    \State Compute $\delta^*$ in \eqref{eq:adp_delta} \Comment{Step size}
    \State $\vx^* \gets \vx^* +\delta^* \delx$ \Comment{Update the solution}
    \State $\vw_i \gets \vw_i + \delta^* (\tilde \vw_i -\vw_i)$ ~ for all $i\in \Gamma$ \Comment{Update $\vw_i$ on the active set}
    \If{$\delta^* < 1$}
        \State $\Gamma \gets \Gamma \backslash \gamma^-$ \Comment{Remove an element from the support}
    \Else
        \State $\gamma^+ = \argmax_{i\in \Gamma^c}{|\va_i^T(\mA\vx^*-\vy)|}$ \Comment{Select the largest among inactive constraints}
       \State $\Gamma \gets \Gamma \cup \gamma^+$ \Comment{Add a new element to the support}
    \EndIf
    \State $\vw_i \gets \max_j|\va_j^T(\mA\vx^*-\vy)|$ ~ for all $i \in \Gamma^c$ \Comment{Update $\vw_i$ on the inactive set}
    \Until{$\max_i{(\vw_i)} \le \tau$}
  \end{algorithmic}
\end{algorithm}

\subsubsection{Weight selection criteria}
Suppose we want to shrink the $\vw_i$ in \eqref{eq:wtBPDN} toward a preset threshold $\tau$, and by construction, we want the $\vw_i$ to have smaller values on the support of the solution (e.g., of the form \eqref{eq:wt_update}). At every homotopy step, we divide the indices into an active and an inactive set. We can follow a number of heuristics to select the desired values of the weights $(\tilde \vw_i)$ so that the $\vw_i$ reduce at a faster rate on the active set than on the inactive set.

{\it Initialization:} We initialize all the weights with a large value (e.g., $\vw_i  = \max_i |\va_i^T\vy| $ for all $i$) for which the solution is a zero vector. The only element in the active set correspond to $\argmax_i|\va_i^T \vy|$, where $\va_i$ denotes $i$th column in $\mA$.

{\it Weights on the active set: }  We can select the $\tilde \vw_i$ on the active set in a variety of ways. For instance, we can select each $\tilde \vw_i$ as a fraction of the present value of the corresponding $\vw_i$ as $\tilde \vw_i \gets \vw_i/\beta$, for some $\beta>1$, or as a fraction of the maximum value of the $\vw_i$ on the active set as $\tilde \vw_i \gets {\max_{i\in \Gamma}{\vw_i}}/{\beta}$. The former will reduce each $\vw_i$ on the active set at the same rate, while the latter will reduce each $\vw_i$ to the same value as well. To introduce reweighting of the form in \eqref{eq:wt_update}, we can change the $\vw_i$ on the support using the solution from the previous homotopy step as $\tilde \vw_i \gets \min\left(\tau,{\tau}/{\beta|\vx^*_i|}\right)$, for some $\beta > 1$.

{\it Weights on the inactive set: } We can assign the $\tilde \vw_i$ on the inactive set a single value that is either equal to the maximum value of $\tilde \vw_i$ on the active set or equal to $\tau$, whichever is the larger.

\begin{figure*}
\centering
    %
    \begin{subfigure}[t]{0.32\textwidth}
        \centering
        \includegraphics[width=\textwidth]{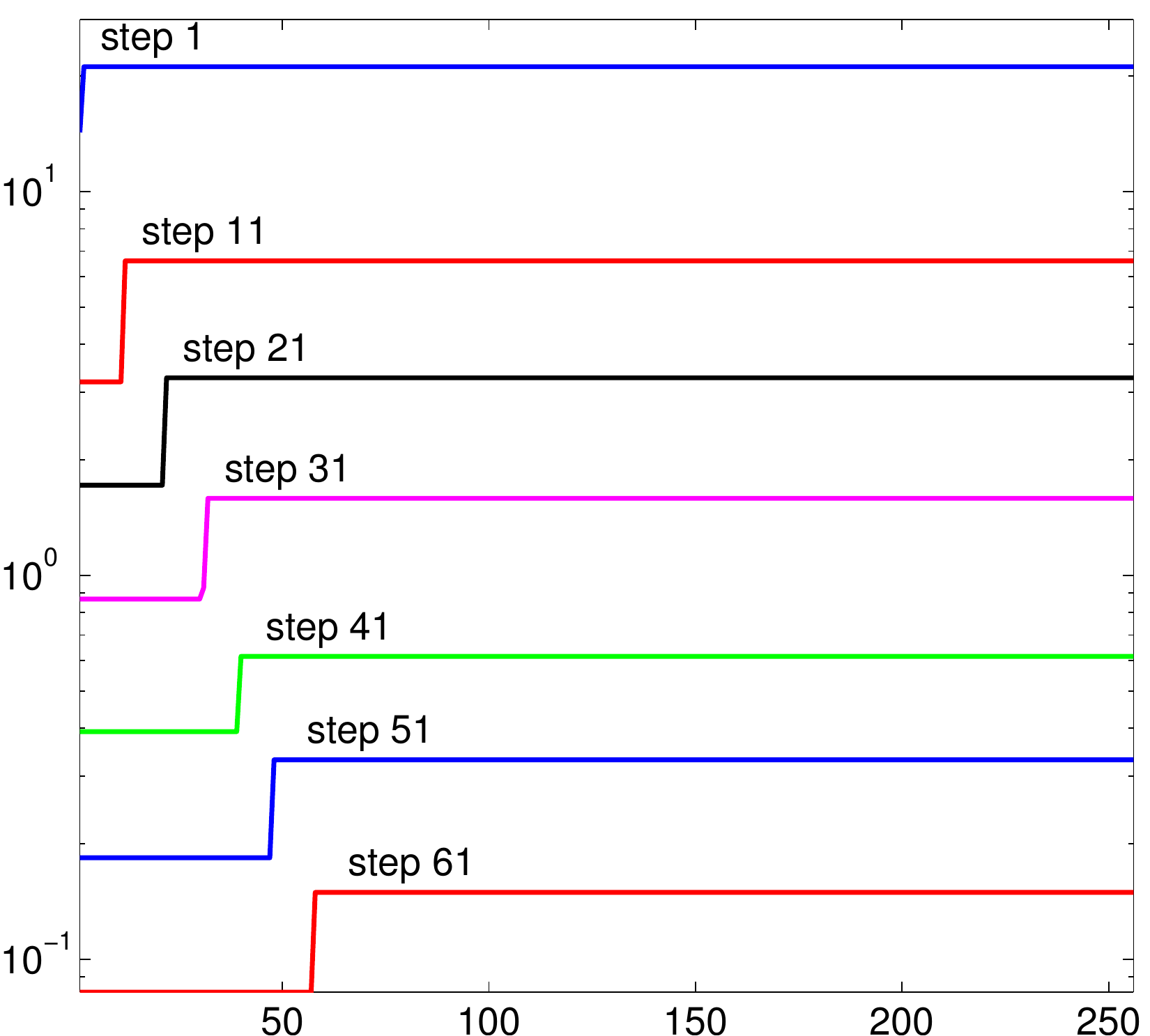}
        \captionsetup{font=footnotesize}
        \caption{$\tilde \vw_i\gets \dfrac{\max_{i\in\Gamma}\vw_i}{2}$}
        \label{fig:Tsteps_evolution}
    \end{subfigure}
    \begin{subfigure}[t]{0.32\textwidth}
        \centering
        \includegraphics[width=\textwidth]{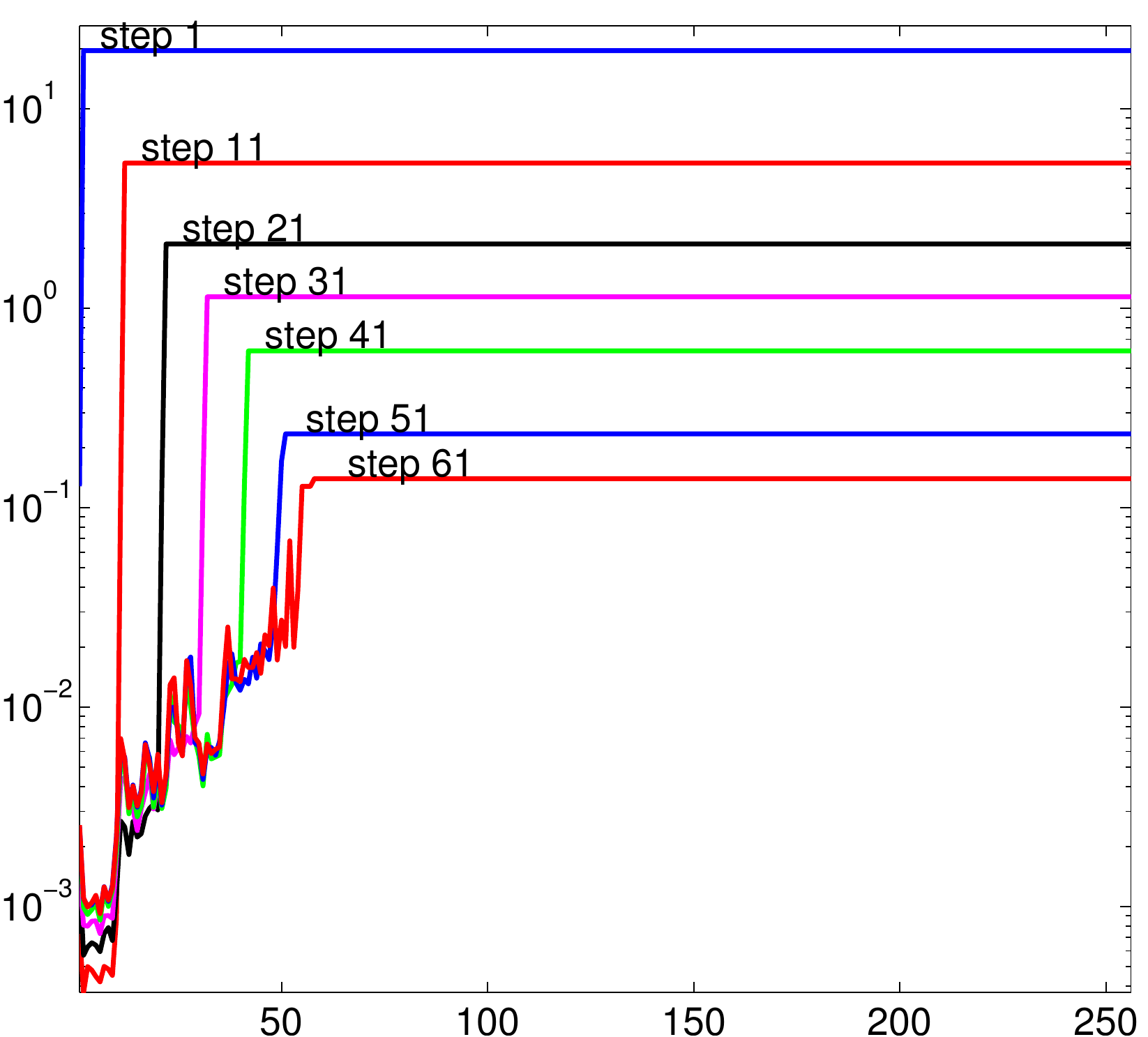}
        \captionsetup{font=footnotesize}
        \caption{$\tilde \vw_i\gets \min\left(\tau,\dfrac{\tau}{\beta |\vx^*_i|}\right)$}
        \label{fig:Trwt_evolution}
    \end{subfigure}
    \begin{subfigure}[t]{0.32\textwidth}
        \centering
        \includegraphics[width=\textwidth]{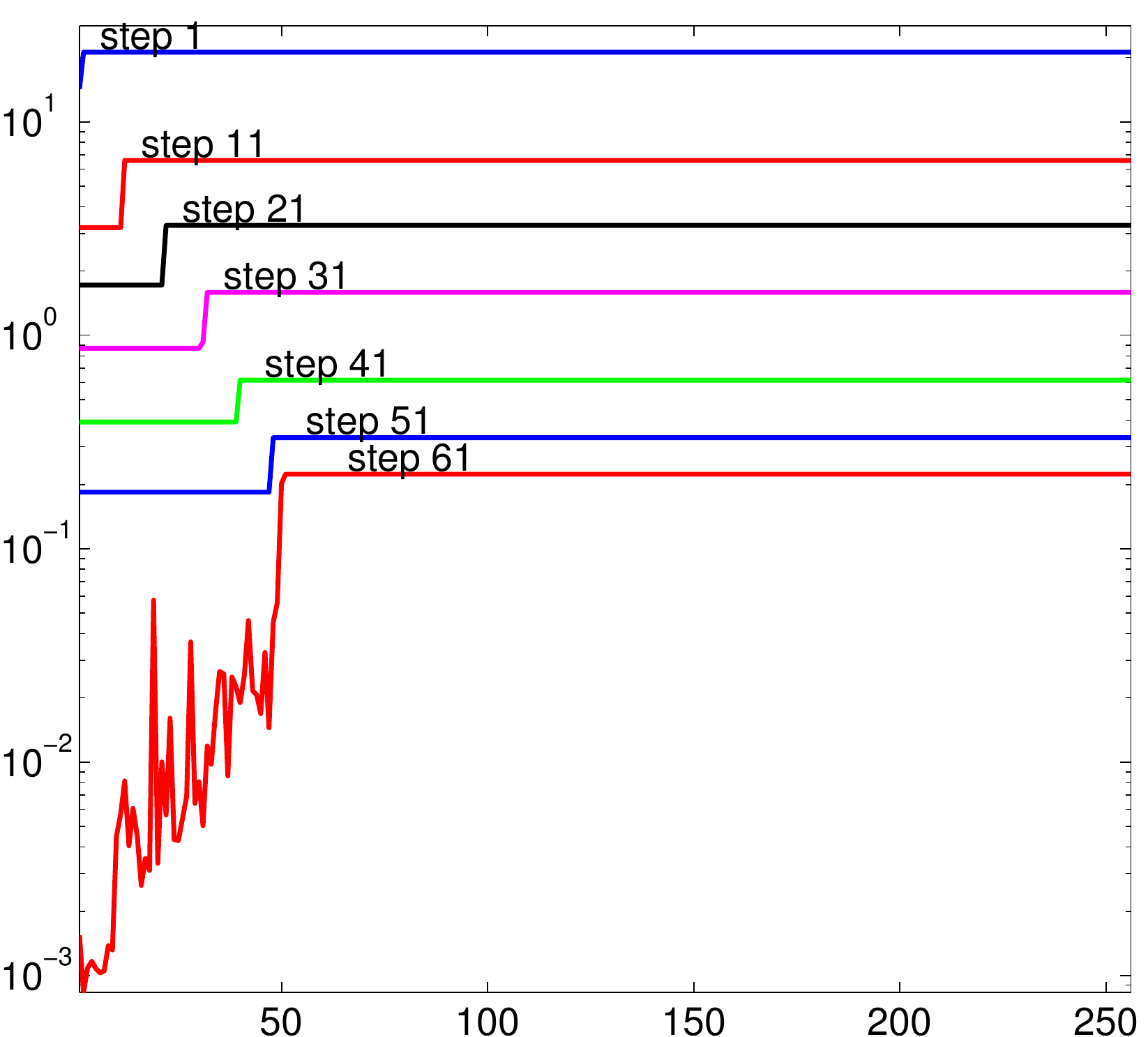}
        \captionsetup{font=footnotesize}
        \caption{a hybrid of (a) and (b)}
        \label{fig:rwt_evolution}
    \end{subfigure}%
    ~ 
    ~ 
    \caption{Illustrations of variations (on a log-scale) in the $\vw_i$ on the sets of active and inactive indices at different homotopy steps. Subfigures (a), (b), and (c) correspond to three different choices for the $\tilde \vw_i$. Left part of each plot (with the lower values of $\vw_i$) corresponds to the active set of indices in the order in which they entered the support and the right part (with larger values of the $\vw_i$) corresponds to the inactive set of the solution at every step.}\label{fig:wt_evolution}
\end{figure*}

In Fig.~\ref{fig:wt_evolution}, we present three examples to illustrate the evolution of the $\vw_i$ on the active and the inactive set at various homotopy steps. These examples were constructed with different choices of $\tilde \vw_i$ during the recovery a Blocks signal of length $256$ from $85$ noisy Gaussian measurements according to the experimental setup described in Sec.~\ref{sec:exp}. We plotted the $\vw_i$ at different homotopy steps in such a way that the left part of each plot corresponds  to the active set and the right part to the inactive set of the solution. As the homotopy progresses, the support size increases and the $\vw_i$ decrease, but the $\vw_i$ on the active set become distinctly smaller than the rest.
%
In Fig.~\ref{fig:Tsteps_evolution} we selected $\tilde \vw_i\gets \max_{i\in \Gamma} \vw_i/2$ at every step;
in Fig.~\ref{fig:Trwt_evolution} we selected $\tilde \vw_i \gets \min\left(\tau,{\tau}/{\beta|\vx^*_i|}\right)$ at every step, with certain values of $\tau$ and $\beta$;
and in Fig.~\ref{fig:rwt_evolution} we selected $\tilde \vw_i \gets {\max_{i\in\Gamma} \vw_i}/{2}$ for first few homotopy steps and then we selected $\tilde \vw_i \gets  {\tau}/{\beta|\vx^*_i|}$. In our experiments in Sec.~\ref{sec:exp}, we selected weights according to the scheme illustrated in Fig.~\ref{fig:Trwt_evolution}.

\subsubsection{Update direction}
To compute the update direction in which the solution moves as we change the weights $\vw_i$ toward the $\tilde \vw_i$, we use the same methodology that we used for \eqref{eq:rwtBPDN} in Sec.~\ref{sec:rwt}. For any given values of the $\vw_i$, a solution $\vx^*$ for \eqref{eq:wtBPDN} satisfies the following optimality conditions:
\begin{subequations}\label{eq:adpBPDN_opt}
\begin{flalign}
&& \va_i^T(\mA\vx^*-\vy) = - \vw_i \vz_i, ~ \text{for all } i \in \Gamma, && \label{eq:adpBPDN_opt1}\\
\text{and} &&|\va_i^T(\mA\vx^*-\vy)| < \vw_i, ~ \text{for all } i \in \Gamma^c,&& \label{eq:adpBPDN_opt2}
\end{flalign}
\end{subequations}
where $\Gamma$ denotes the support of $\vx^*$ and $\vz$ denotes the sign sequence of $\vx^*$ on $\Gamma$.
If we change $\vw_i$ toward $\tilde \vw_i$ along a straight line, $(1-\delta)\vw_i+\delta \tilde \vw_i$, the solution moves in a direction $\delx$, which to maintain optimality must obey
\begin{subequations}\label{eq:adpBPDN_opt}
\begin{gather}
\mA_\Gamma^T(\mA\vx^*-\vy)+ \delta\mA_\Gamma^T\mA \delx = - \mW \vz + \delta (\mW-\widetilde \mW)\vz, \label{eq:adpBPDN_optUpdate1}\\
|\va_i^T(\mA\vx^*-\vy)+ \delta \va_i^T\mA\delx| < \vw_i + \delta(\tilde \vw_i - \vw_i), ~\text{for all } i \in \Gamma^c, \label{eq:adpBPDN_optUpdate2}
\end{gather}
\end{subequations}
where $\mW$ and $\widetilde \mW$ denote $|\Gamma|\times|\Gamma|$ diagonal matrices constructed with the respective values of $\vw_i$ and $\tilde \vw_i$ on $\Gamma$. Subtracting \eqref{eq:adpBPDN_opt1} from \eqref{eq:adpBPDN_optUpdate1} yields the following expression for the update direction $\delx$:
\begin{equation}\label{eq:adp_delx}
\delx =
\begin{cases}
(\mA_\Gamma^T\mA_\Gamma)^{-1} (\mW-\widetilde \mW)\vz & \text{on } \Gamma \\
0 & \text{on } \Gamma^c.
\end{cases}
\end{equation}

\subsubsection{Step size and support selection}
As we increase $\delta$ from 0 to 1, $\vx^*$ moves along the update direction $\delx$ as $\vx^*+\delta \delx$ and the $\vw_i$ change toward $\tilde \vw_i$ as $\vw_i + \delta (\tilde \vw_i - \vw_i)$. At certain value of $\delta\in(0,1)$, an existing element in $\vx^*$ may shrink to zero and we must remove that element from the support. Alternatively, an inactive constraint in \eqref{eq:adpBPDN_optUpdate2} may become active, and to maintain the optimality of the solution, we must either include that element in the support or increase the value of $\vw_i$ at that index.

%
The optimality conditions \eqref{eq:adpBPDN_opt2} and \eqref{eq:adpBPDN_optUpdate2} suggest that as long as a strict inequality is maintained for an index $i$ in the inactive set, we can change the corresponding weight to an arbitrary value without affecting the solution.
Sine the values of $\vw_i$ are not fixed a priori in this scheme, we have the flexibility to disregard any violation of the inequality constraints and adjust the $\vw_i$ so that the solution remains optimal.
Under this setting, we can compute the optimal stepsize $\delta^*$ and identify a change in the support of the signal as follows.

The smallest positive value of $\delta$ that causes an element in $\vx^*$ to shrink to zero is
\begin{equation}
\delta^- = \min_{i\in \Gamma} \left(\frac{-\vx^*_i}{\delx_i}\right)_+,
\end{equation}
suppose at an index $\gamma^- \in \Gamma$.
If $\delta^- < 1$, we must remove $\gamma^-$ from the support and set $\delta^* = \delta^-$. If $\delta^- > 1$, we set $\delta^* = 1$ and select a new element $\gamma^+$ to add to the support. We set
\begin{equation}\label{eq:adp_delta}
\delta^* = \min{(\delta^-,1)},
\end{equation}
$\vx^* = \vx^* + \delta^*\delx$, and $\vw_i = \vw_i + \delta^* (\tilde \vw_i - \vw_i)$. If $\delta^- > 1$, we select the new element $\gamma^+$ that corresponds to the inactive constraint with largest magnitude, which can be determined as
\begin{equation}\label{eq:adp_gamma+}
\gamma^+ = \argmax_{i\in \Gamma^c} |\va_i^T(\mA\vx^*-\vy)|,
\end{equation}
and set $\vw_{\gamma^+} = |\va_{\gamma^+}^T(\mA\vx^*-\vy)|$. We update the $\vw_i$, $\vx^*$, and $\Gamma$ accordingly.

We repeat the procedure of selecting $\tilde \vw_i$, computing the update direction and the stepsize, and updating the solution and its support at the every homotopy step, until a termination criterion is satisfied (e.g. $\vw_i \le \tau$ for all $i$).

The main computational cost at every homotopy step comes from solving a $|\Gamma|\times |\Gamma|$ system of equations in \eqref{eq:adp_delx} for computing $\delx$ and one matrix-vector multiplication whenever we need to find $\gamma^+$ in \eqref{eq:adp_gamma+}. Since $\Gamma$ changes by a single element at every homotopy step, the update direction can be efficiently computed using a rank-one update. As such, the computational cost of every step is equivalent one matrix-vector multiplication with $\mA$ and one with $\mA^T$.

\section{Numerical experiments}\label{sec:exp}
We present some experiments to demonstrate the performance of our proposed algorithms:
(1) iterative reweighting via homotopy (Algorithm~\ref{alg:rwt}), which we will call IRW-H  and
(2) adaptive reweighting via homotopy (Algorithm~\ref{alg:adp}), which we will call ARW-H.
We evaluate the performances of ARW-H and IRW-H in terms of the computational cost and the reconstruction accuracy.
We show that, in comparison with iterative reweighting schemes, solving \eqref{eq:wtBPDN} using ARW-H yields significantly higher quality signal reconstruction, at a computational cost that is comparable to solving \eqref{eq:wtBPDN} one time from scratch.
Furthermore, we show that, using IRW-H, we can quickly update the weights in \eqref{eq:wtBPDN} at a small computational expense.
To compare ARW-H and IRW-H against existing $\ell_1$ solvers, we also present results for sparse signal recovery using iterative reweighting for three state-of-the-art solvers\footnote{We selected these solvers for comparison because, among the commonly used $\ell_1$ solvers \cite{afonso-2010-salsa,Figueiredo_2007_GPSR,hale2008fpc}, we found these to be the fastest and sufficiently accurate with a warm start.}: YALL1~\cite{yang-2011-yall1}, SpaRSA~\cite{Wright-2009-sparsa}, and SPGL1~\cite{BergFriedlander-2008-spgl1} in which we used old solutions as a ``warm start'' at every iteration of reweighting.
We show that IRW-H outperforms YALL1, SpaRSA, and SPGL1 in terms of the computational cost for iterative reweighting, while ARW-H yields better overall performance in terms of both the computational cost and the reconstruction accuracy.

\subsection{Experimental setup}
\begin{figure*}
\centering
    \begin{subfigure}[t]{0.49\textwidth}
        \includegraphics[page=1,trim = 0mm 0mm 0mm 0mm, clip, width=1\columnwidth,keepaspectratio]{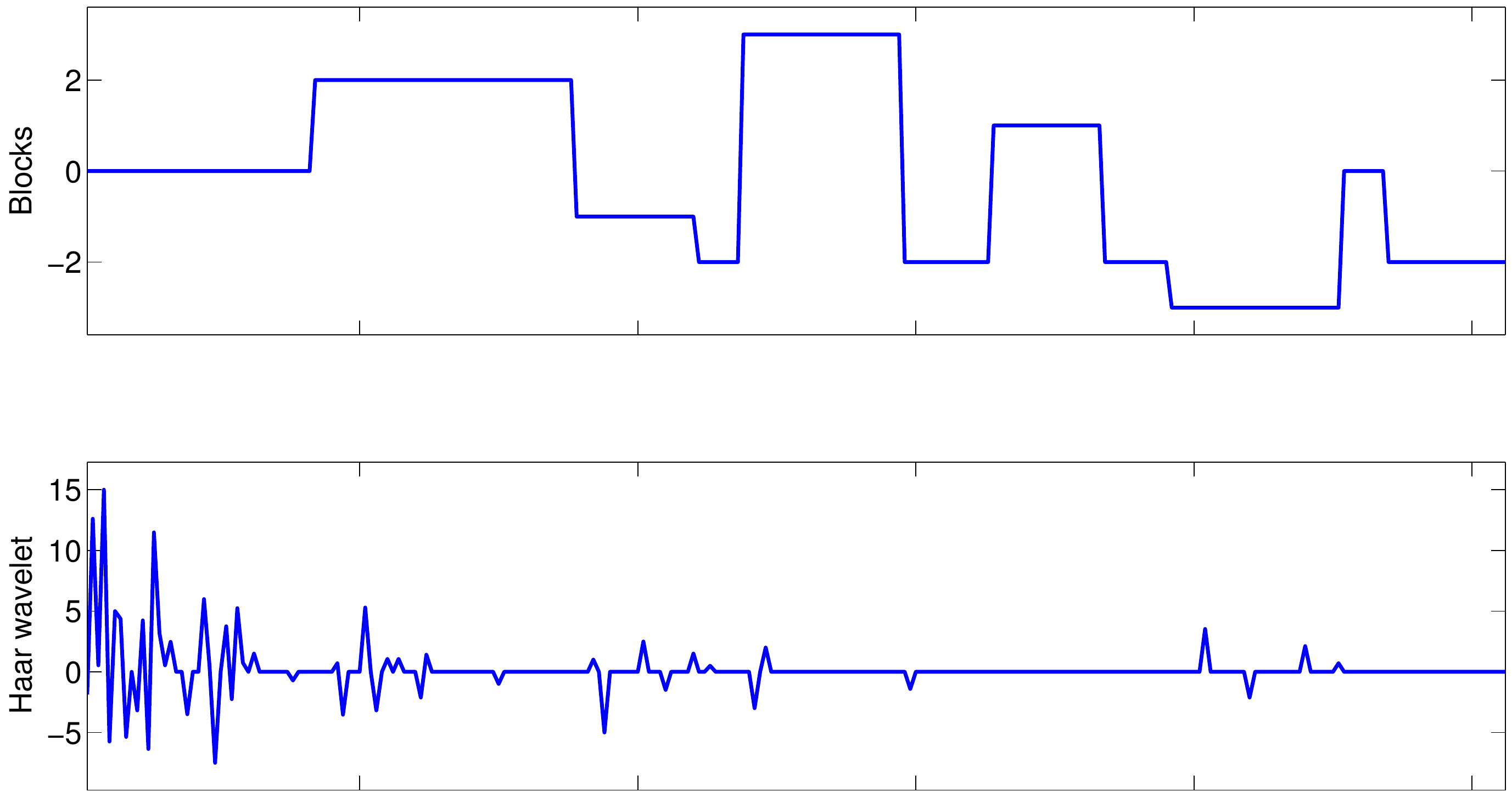}
        \caption{Blocks}
        \label{fig:blocks}
    \end{subfigure}
    \begin{subfigure}[t]{0.49\textwidth}
        \includegraphics[page=1,trim = 0mm 0mm 0mm 0mm, clip, width=1\columnwidth,keepaspectratio]{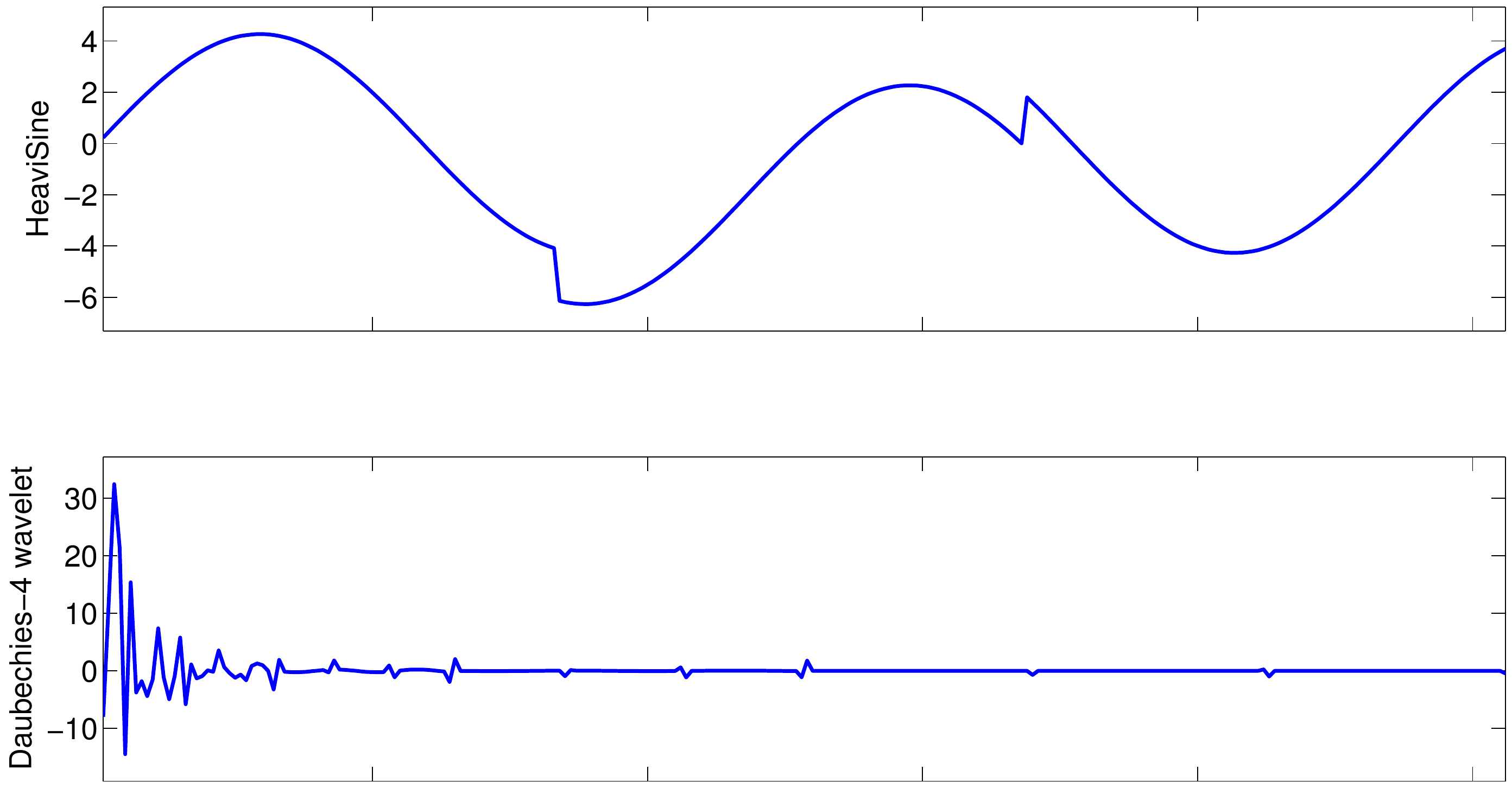}
        \caption{HeaviSine}
        \label{fig:HeaviSine}
    \end{subfigure}
    \caption{{ (a)}~An example of piecewise-constant (blocks) signal and its sparse representation using Haar wavelets. { (b)}~An example of perturbed HeaviSine signal and its sparse representation using Daubechies-4 wavelets.}\label{fig:largescale}
\end{figure*}

We compared the performances of the algorithms above in recovering two types of sparse signals from noisy, random measurements that were simulated according to the model in \eqref{eq:y=Ax+e}.
We generated sparse signals by applying wavelet transforms on the modified forms of ``Blocks'' and ``HeaviSine'' signals from the Wavelab toolbox~\cite{Wavelab} as described below.
\begin{enumerate}[i.]
\item {\bf Blocks}: We generated a piecewise-constant signal of length $N$ by randomly dividing the interval $[1,~N]$ into 11 disjoint regions. Setting the first region to zero, we iteratively assigned a random value to every region by adding an integer chosen uniformly in the range of $[-5,~5]$ to the value from the previous region. We applied Haar wavelet transform on the piecewise-constant signal to generate a sparse signal $\bar\vx$. An example of such a piecewise-constant signal and its Haar wavelet transform is presented in Fig.~\ref{fig:blocks}.
    Because of the piecewise constant structure of these signals, the resulting Haar wavelet transforms will have only a small number of nonzero coefficients that depend on the number of discontinuities and the finest wavelet scale. Since we have fixed the number of discontinuities, the ratio of the number of nonzero elements to the length of the signal becomes smaller as the length of the signal ($N$) increases.
\item {\bf HeaviSine}: We generated a sinusoidal signal with nearly two cycles and two jumps at random locations. First, we generated a sinusoidal signal of length $N$ for which we selected the amplitude in the range of $[4,~6]$ and number of cycles in the range $[2,~2.5]$ uniformly at random. Then, we divided the signal into three non-overlapping regions and added a different Gaussian random variable to each region. We applied Daubechies~4 wavelet transform on the resulting signal to generate the sparse signal $\bar\vx$. An example of such a sinusoidal signal with jumps and its Daubechies~4 wavelet transform is presented in Fig.~\ref{fig:HeaviSine}.
    In this type of signals, most of the wavelet coefficients in $\bar\vx$ will not be exactly zero, but if sorted in the decreasing order of magnitude, the coefficients quickly decay to extremely small values. Hence, this type of signals can be classified as near-sparse or compressible.
\end{enumerate}
In every experiment, we generated an $M\times N$ measurement matrix $\mA$ with its entries drawn independently according to $\mathcal{N}(0,1/\sqrt{M})$ distribution and added Gaussian noise vector $\ve$ to generate the measurement vector as $\vy = \mA\bar\vx+\ve$. We selected each entry in $\ve$ as i.i.d. $\mathcal{N}(0,\sigma^2)$, where the variance $\sigma^2$ was selected to set the expected SNR with respect to the measurements $\mA\bar\vx$ at 40\,dB.
We reconstructed solution $\widehat \vx$ using all the algorithms according to the procedures described below.

In our experiments, we fixed the parameter $\tau = \sigma \sqrt{\log N}$, where $\sigma$ denotes the standard deviation of the measurement noise. Although the weights can be tuned according to the signal structure, measurement matrix, and noise level, we did not make such an attempt in our comparison. Instead, we adopted a general rule for selecting weights that provided good overall performance for all the solvers, in all of our experiments.
We set up the algorithms in the following manner.
\begin{enumerate}[i.]
\item {\bf ARW-H\footnote{\label{fn:L1-homotopy}MATLAB code added to the $\ell_1$-homotopy package available at http://users.ece.gatech.edu/$\sim$sasif/homotopy. Additional experiments and scripts to reproduce all the experiments in this paper can also be found on this webpage.}}: We solved a weighted $\ell_1$-norm minimization problem of the form \eqref{eq:wtBPDN} following the procedure outlined in Algorithm~\ref{alg:adp}, in which the exact values for the $\vw_i$ are not known a priori as they are selected adaptively. In line \ref{alg:adp_wtUpdate} of Algorithm~\ref{alg:adp}, we selected
    \mbox{$\tilde \vw_i \gets \min\left(\tau,\dfrac{\tau}{\beta |\vx^*_i|}\right)$} using \mbox{$\beta \gets M\dfrac{\|\vx^*\|_2^2}{\|\vx^*\|_1^2}$}
    at every step, where $\vx^*$ denotes the solution from the previous homotopy step.
    We selected this value of $\beta$ because it helps in shrinking the $\vw_i$ to smaller values when $M$ is large and to larger values when the solution is dense. (We are using the ratio of $\ell_1$ to $\ell_2$ norm as a proxy for the support size here.)
    The main computational cost at every step of ARW-H involves one matrix-vector multiplication for identifying a change in the support and a rank-one update for computing the update direction. We used the matrix inversion lemma-based scheme to perform the rank-one updates.
    %
    %
\item  {\bf IRW-H\footref{fn:L1-homotopy}}: We solved \eqref{eq:wtBPDN} via iterative reweighting in which we updated the solution at every reweighting iteration according to the procedure outlined in Algorithm~\ref{alg:rwt}. For the first iteration, we used standard LASSO homotopy algorithm~\cite{AR_l1_homotopy_webpage} to solve \eqref{eq:BPDN}, which is equivalent to \eqref{eq:wtBPDN} when $\vw_i = \tau$ for all $i$.
    Afterwards, at every reweighting iteration, we updated the $\vw_i$ as
    \begin{equation}\label{eq:update_wt}
    \vw_i \gets \frac{\tau}{\beta|\widehat \vx_i| + \epsilon},
    \end{equation}
    where $\widehat{\vx}$ denotes the solution from previous reweighting iteration and $\beta\ge1$ and $\epsilon>0$ denote two parameters that can be used to tune the weights according to the problem. In our experiments, we fixed $\epsilon=1$ and updated $\beta \gets M\dfrac{\|\widehat \vx\|_2^2}{\|\widehat \vx\|_1^2}$ at every reweighting iteration. The main computational cost at every step of IRW-H also involves one matrix-vector multiplication and a rank-one update of a small matrix. We used matrix inversion lemma-based scheme to perform rank-one updates.
\item {\bf YALL1\footnote{MATLAB package for YALL1 is available at http://yall1.blogs.rice.edu/.}}: YALL1 is a first-order algorithm that uses an alternating direction minimization method for solving various $\ell_1$ problems, see~\cite{yang-2011-yall1} for further details. We iteratively solved \eqref{eq:wtBPDN} using weighted-$\ell_1/\ell_2$ solver in YALL1 package. For the initial solution, we solved \eqref{eq:BPDN} using YALL1. At every subsequent reweighting iteration, we used previous YALL1 solution to renew the weights according to \eqref{eq:update_wt} and solved \eqref{eq:wtBPDN} by providing the old solution as a warm-start to YALL1 solver. We fixed the tolerance parameter to $10^{-4}$ in all the experiments. The main computational cost of every step in the YALL1 solver comes from applications of $\mA$ and $\mA^T$.
\item {\bf SpaRSA\footnote{MATLAB package for SpaRSA is available at http://lx.it.pt/$\sim$mtf/SpaRSA/.}}: SpaRSA is also a first-order method that uses a fast variant of iterative shrinkage and thresholding for solving various $\ell_1$-regularized problems, see~\cite{Wright-2009-sparsa} for further details. Similar to IRW-H and YALL1, we iteratively solved \eqref{eq:wtBPDN} using SpaRSA, while updating weights using the old solution in \eqref{eq:update_wt} and using the old solution as a warm-start at every reweighting iteration.
    We used the SpaRSA code with default adaptive continuation procedure in the Safeguard mode using the duality gap-based termination criterion for which we fixed the tolerance parameter to $10^{-4}$ and modified the code to accommodate weights in the evaluation.
    The main computational cost for every step in the SpaRSA solver also involves applications of $\mA$ and $\mA^T$.
\item {\bf SPGL1\footnote{MATLAB package for SPGL1 is available at http://www.cs.ubc.ca/labs/scl/spgl1}}: SPGL1 solves an equivalent constrained form of \eqref{eq:wtBPDN} by employing a root-finding algorithm~\cite{BergFriedlander-2008-spgl1}. We solved the following problem using SPGL1:
    \begin{equation}\label{eq:wtBPDN_spgl1}
    \underset{\vx}{\text{minimize}} \; \sum_{i=1}^N \vw_i |\vx_i|~\text{subject to}~ \|\mA\vx -\vy\|_2\le \lambda,
    \end{equation}
    in which we used $\lambda=\sigma \sqrt{M}$. For the initial solution, we solved \eqref{eq:wtBPDN_spgl1} using $\vw_i = 1$ for all $i$. At every subsequent reweighting iteration, we used previous SPGL1 solution to renew the weights according to \eqref{eq:update_wt} (using $\tau = 1$) and solved \eqref{eq:wtBPDN_spgl1} using the old solution as a warm start. We solved SPGL1 using default parameters with optimality tolerance set at $10^{-4}$. The computational cost of every step in SPGL1 is also dominated by matrix-vector multiplications.
\end{enumerate}
To summarize, ARW-H solves \eqref{eq:wtBPDN} by adaptively selecting the values of $\vw_i$, while IRW-H, YALL1, and SpaRSA iteratively solve \eqref{eq:wtBPDN} and SPGL1 iteratively solves \eqref{eq:wtBPDN_spgl1}, using updated values of $\vw_i$ at every reweighting iteration.

We used MATLAB implementations of all the algorithms and performed all the experiments on a standard desktop computer using MATLAB~2012a. We used a single computational thread for all the experiments, which involved recovery of a sparse signal from a given set of measurements using all the candidate algorithms. In every experiment, we recorded three quantities for each algorithm: 1) the quality of reconstructed signal in terms of signal-to-error ratio in dB, defined as
$$ \text{SER} = 20\log_{10}\frac{\|\bar\vx\|_2}{\|\bar\vx-\widehat\vx\|_2},$$
where $\bar\vx$ and $\widehat \vx$ denote the original and the reconstructed signal, respectively, 2) the number of matrix-vector products with $\mA$ and $\mA^T$, and 3) the execution time in MATLAB.

\subsection{Results}
We compared performances of ARW-H, IRW-H, YALL1, SpaRSA, and SPGL1 for the recovery of randomly perturbed Blocks and HeaviSine signals from random, noisy measurements.
We performed 100 independent trials for each of the following combinations of $N$ and $M$: $N = [256,~512,~1024]$ and $M = [N/2,~ N/2.5,~N/3,~N/3.5,~N/4]$. In each experiment, we recovered a solution $\widehat \vx$ from simulated noisy, random measurements using all the algorithms, according to the procedures described above, and recorded corresponding SER, number of matrix-vector products, and MATLAB runtime. The results, averaged over all the trials, for each combination of $M$ and $N$ are presented in Figures~\ref{fig:sType-blocks_SNR}--\ref{fig:sType-blocks_CPU} (for Blocks signals) and Figures~\ref{fig:sType-HeaviSine_SNR}--\ref{fig:sType-HeaviSine_CPU} (for HeaviSine signals).

Comparison of SERs for the solutions of all the algorithms at different values of $N$ and $M$ is presented in Fig.~\ref{fig:sType-blocks_SNR} (for Blocks signals) and Fig.~\ref{fig:sType-HeaviSine_SNR} (for HeaviSine signals). Three plots in the first row depict SERs for the solutions after first iteration of all the algorithms. Since ARW-H solves a weighted $\ell_1$-norm formulation (as in \eqref{eq:wtBPDN}) via adaptive reweighting, its performance is superior to all the other algorithms, which solve unweighted $\ell_1$-norm problems in their first iteration.
Since SPGL1 solves the $\ell_1$ problem in \eqref{eq:wtBPDN_spgl1}, its performance is slightly different compared to IRW-H, YALL1, and SpaRSA, all of which solve \eqref{eq:wtBPDN} and should provide identical solutions if they converge properly.
The plots in the second row present SERs for the solutions after five reweighting iterations of all the algorithms except ARW-H, which was solved only once.
As we can see that the solutions of ARW-H display the best SERs in all these experiments.
Although SERs for the solutions of IRW-H, YALL1, SpaRSA, and SPGL1 improve with iterative reweighting, in some cases there is a significant gap between their SERs and that of ARW-H.


\newcommand{\figwidth}{0.45\textwidth}
\newif\iffullfigure

\fullfiguretrue
\fullfigurefalse

\iffullfigure
\begin{figure}[t]
    \centering
    \includegraphics[page=4,trim = 0mm 0mm 0mm 0mm, clip,  width=1\columnwidth, keepaspectratio]{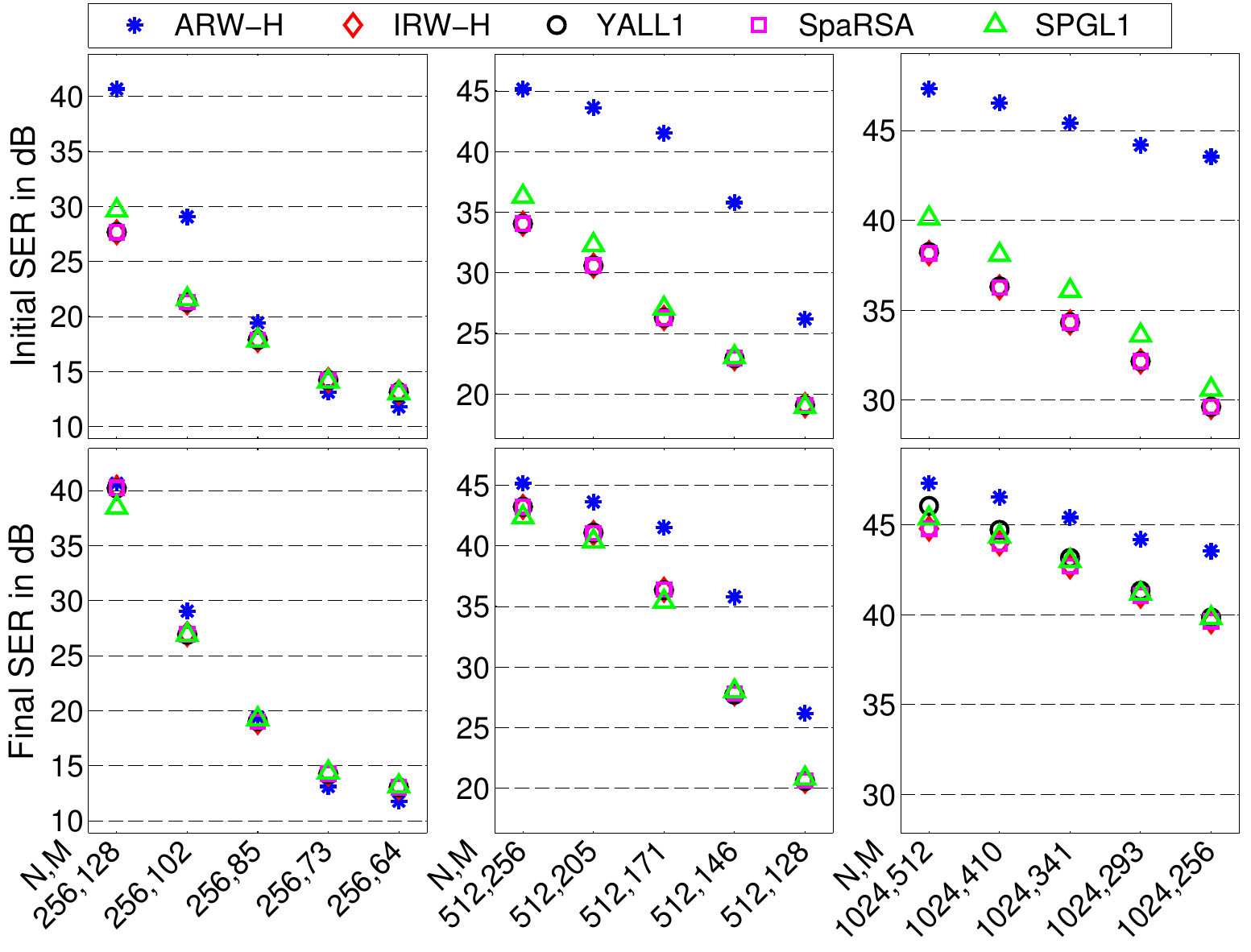}
    \caption{Comparison of SER for the recovery of sparse signals that were constructed by taking Haar wavelet transform of randomly perturbed ``Blocks'' signals and measured with $M\times N$ Gaussian matrices in the presence of Gaussian measurement noise at 40\,dB SNR.
    ARW-H solves adaptive-reweighted $\ell_1$ problem once, while other methods solve unweighted $\ell_1$ problem in their first iteration and perform five reweighted iterations afterwards.
    {\bf (First row)} SER for the solution after first iteration. {\bf (Second row)} SER for solutions after five reweighting iterations (SER for ARW-H is copied from the top row)}
    \label{fig:sType-blocks_SNR}
\end{figure}

\begin{figure}[t]
    \centering
    \includegraphics[page=13,trim = 0mm 0mm 0mm 0mm, clip,  width=1\columnwidth, keepaspectratio]{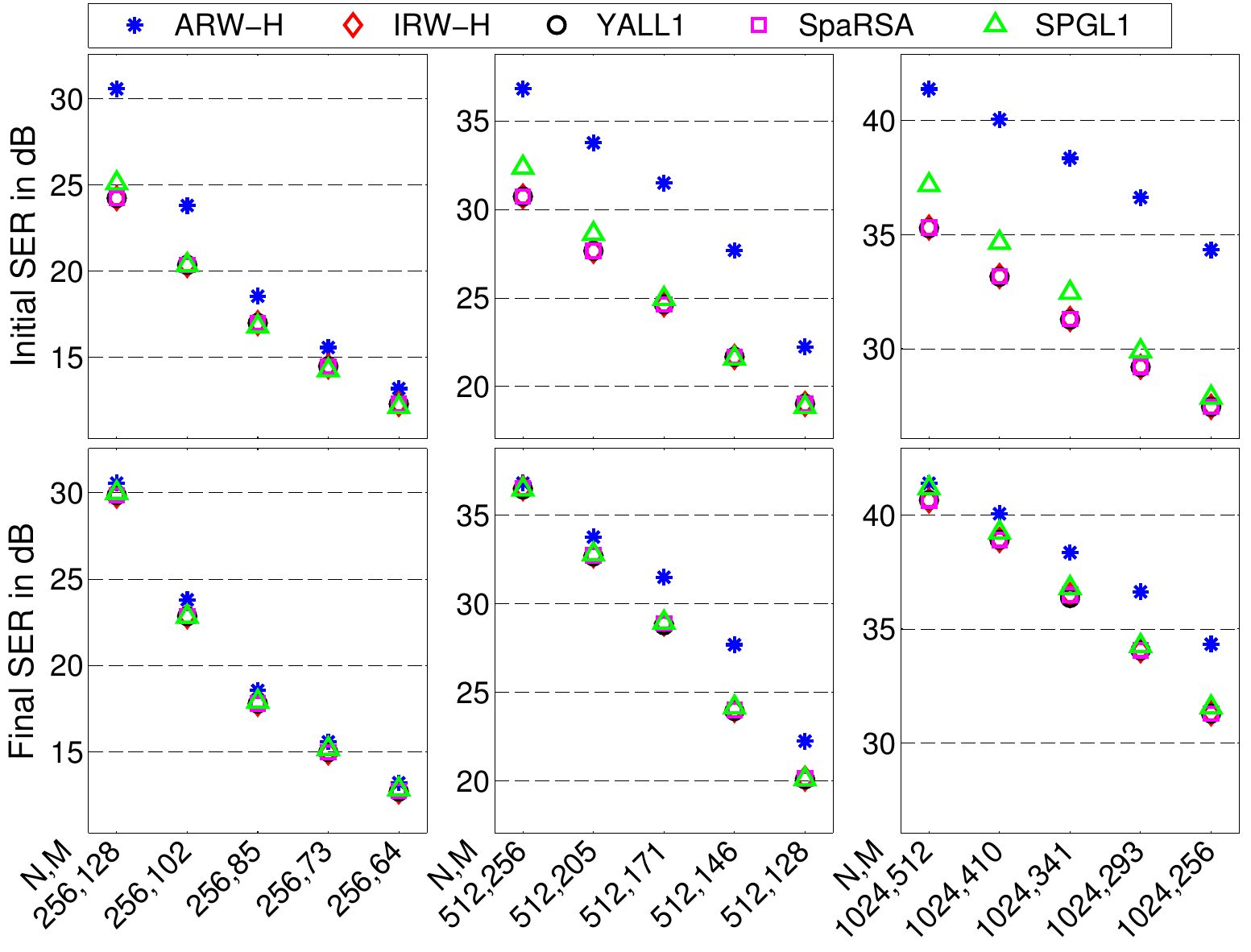}
    \caption{Comparison of SER for the recovery of near-sparse signals that were constructed by taking Daubechies~4 wavelet transform of randomly perturbed HeaviSine signals and measured with $M\times N$ Gaussian matrices in the presence of Gaussian noise at 40\,dB SNR.
    ARW-H solves adaptive-reweighted $\ell_1$ problem once, while other methods solve unweighted $\ell_1$ problem in their first iteration and perform five reweighted iterations afterwards.
    {\bf (First row)} SER for the solution after first iteration. {\bf (Second row)} SER for solutions after five reweighting iterations (SER for ARW-H is copied from the top row)}
    \label{fig:sType-HeaviSine_SNR}
\end{figure}

\else

\begin{figure*}
\hspace{-35pt}
      \begin{subfigure}[t]{\figwidth}
        \centering
        \includegraphics[page=4,trim = 0mm 0mm 0mm 0mm, clip,  width=1.2\textwidth, keepaspectratio]{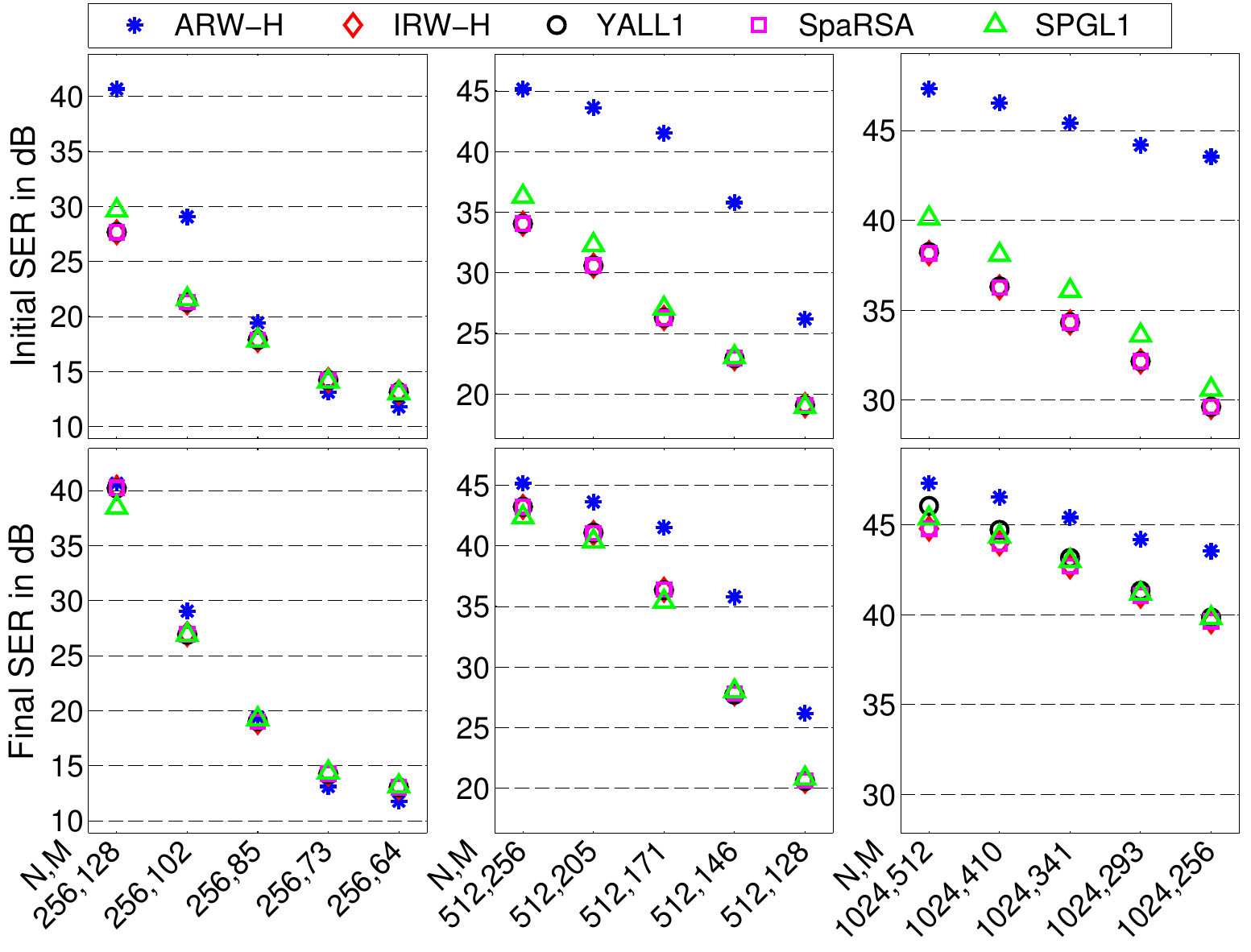}
        \caption{SER --- Blocks}
    \label{fig:sType-blocks_SNR}
    \end{subfigure}
    \hspace{40pt}
    \begin{subfigure}[t]{\figwidth}
        \centering
        \includegraphics[page=13,trim = 0mm 0mm 0mm 0mm, clip, width=1.2\textwidth, keepaspectratio]{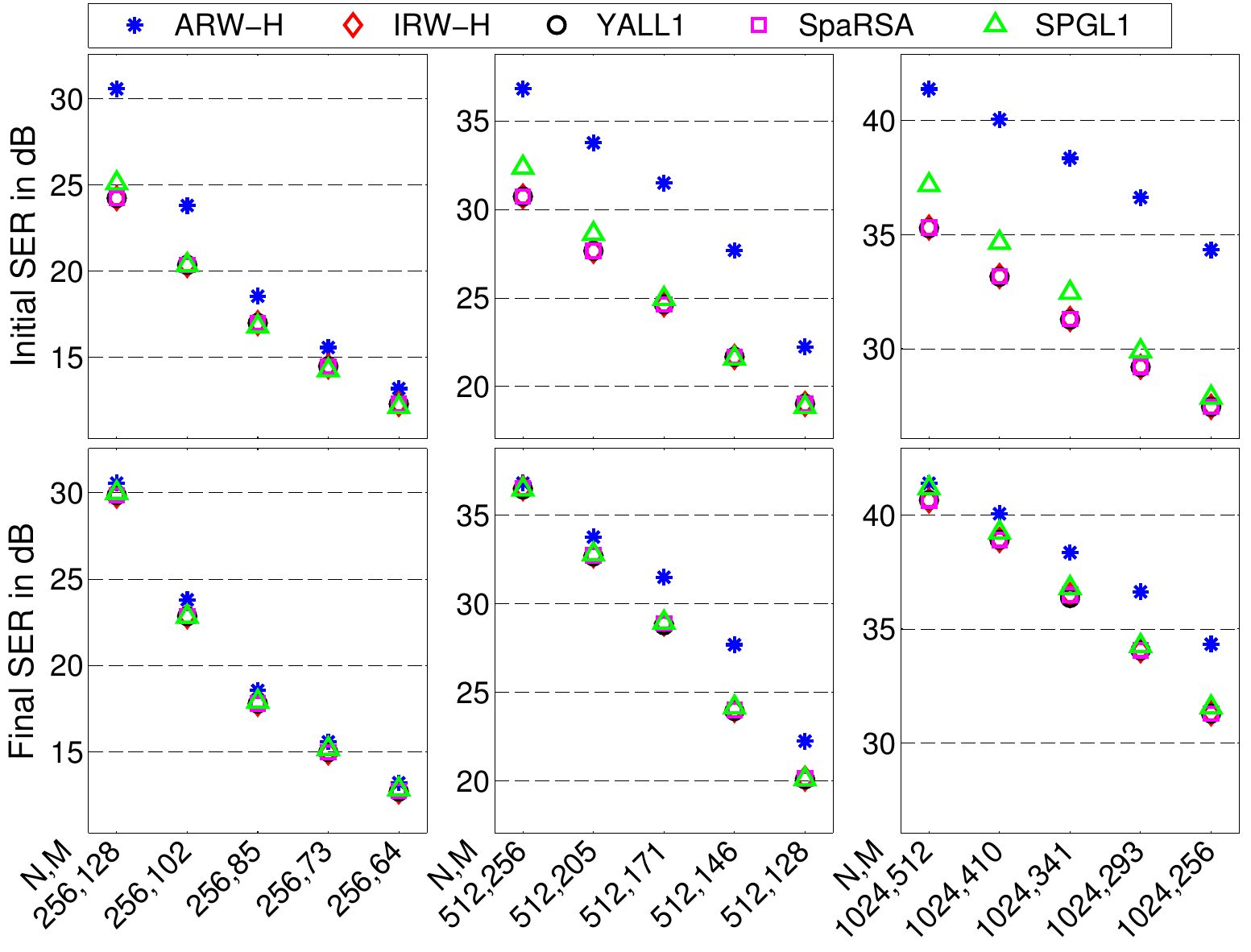}
        \caption{SER --- HeaviSine}
        \label{fig:sType-HeaviSine_SNR}
    \end{subfigure}
    ~ 
    ~ 
    \caption{Comparison of SER for the recovery of sparse signals: ({\bf Left}) Haar wavelet transform of randomly perturbed Blocks signals and {\bf (Right)} Daubechies~4 wavelet transform of randomly perturbed HeaviSine signals, measured with $M\times N$ Gaussian matrices in the presence of Gaussian noise at 40\,dB SNR.
    ARW-H solves adaptive-reweighted $\ell_1$ problem once, while other methods solve unweighted $\ell_1$ problem in their first iteration and perform five reweighted iterations afterwards.
    {\bf (First row)} SER for the solution after first iteration. {\bf (Second row)} SER for solutions after five reweighting iterations (SER for ARW-H is copied from the top row)}
    \label{fig:AtA}
\end{figure*}

\fi

Comparison of the computational cost of all the algorithms in terms of the number of matrix-vector multiplications is presented in Fig.~\ref{fig:sType-blocks_AtA} (for Blocks signals) and Fig.~\ref{fig:sType-HeaviSine_AtA} (for HeaviSine signals). We counted an application of each $\mA$ and $\mA^T$ as one application of $\mA^T\mA$\footnote{For the homotopy algorithms, we approximated the cost of one step as one application of $\mA^T\mA$.}.
Three plots in the first row present the count of $\mA^T\mA$ applications that each algorithm used for computing the initial solution. Second row depicts the count of $\mA^T\mA$ applications summed over five reweighting iterations in each of the algorithm. Since we solved ARW-H just once, the count for ARW-H is zero and does not appear in the second row. Third row presents total count of $\mA^T\mA$ applications, which is the sum of the counts in the first and the second row. We can see in the second row that, compared to YALL1, SPGL1, and SpaRSA, IRW-H used a distinctly smaller number of matrix-vector products for updating the solution as the weights change in iterative reweighting. The final count in the third row shows that ARW-H consumed the least number of total $\mA^T\mA$ applications in all the cases.

\iffullfigure
\begin{figure}[t]
    \centering
    \includegraphics[page=5,trim = 0mm 0mm 0mm 0mm, clip,  width=1\columnwidth, keepaspectratio]{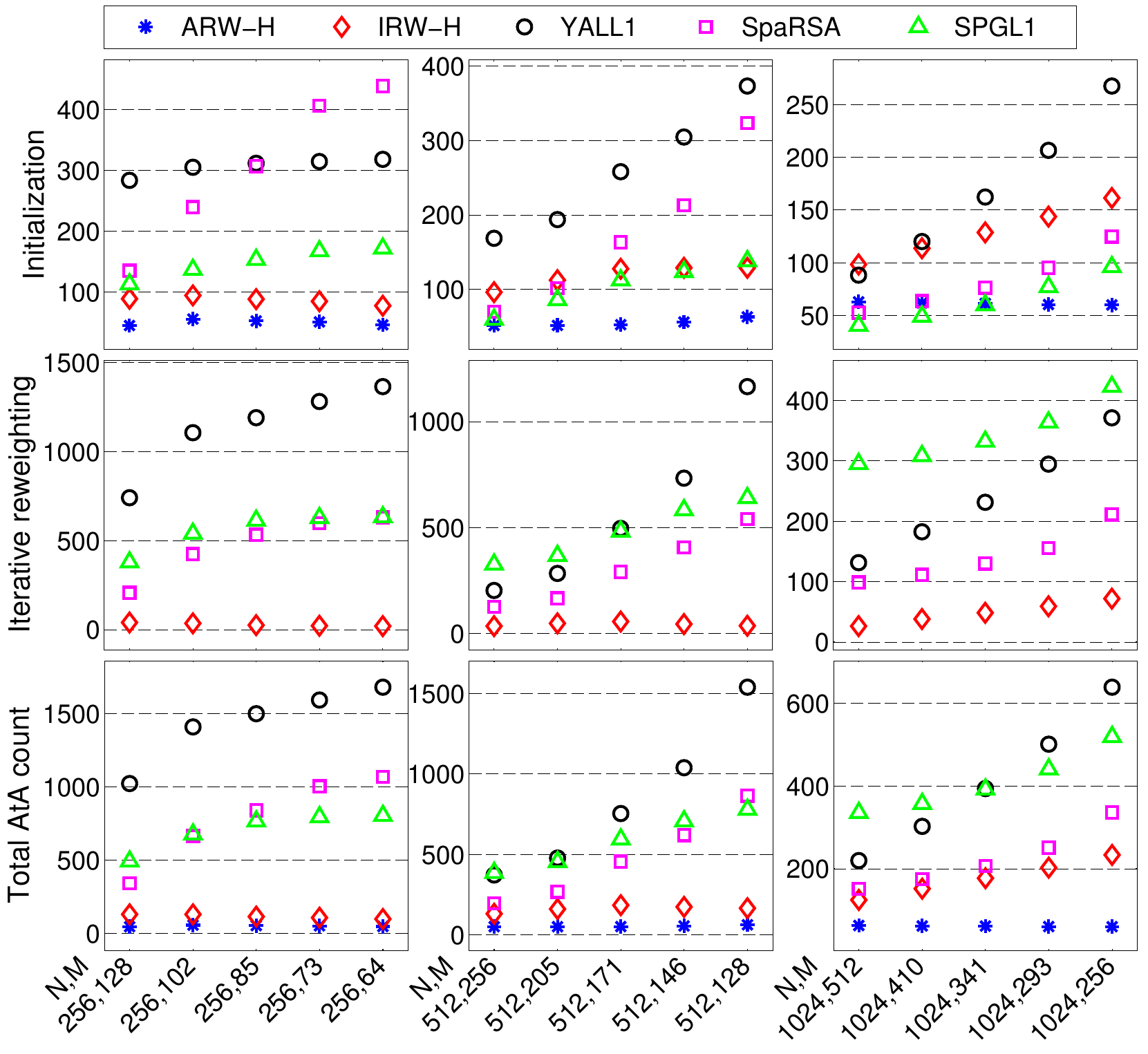}
    \caption{Comparison of the number of matrix-vector products for the recovery of Blocks signals.
    ARW-H solves adaptive-reweighted $\ell_1$ problem once, while other methods solve unweighted $\ell_1$ problem in their first iteration and perform five reweighted iterations afterwards.
    {\bf (First row)} Count for the first iteration only. {\bf (Second row)} Count for all the reweighting iterations (ARW-H does not appear because its count is zero). {\bf (Third row)} Count for all the iterations.}
    \label{fig:sType-blocks_AtA}
\end{figure}
\begin{figure}[t]
    \centering
    \includegraphics[page=14,trim = 0mm 0mm 0mm 0mm, clip, width=1\columnwidth, keepaspectratio]{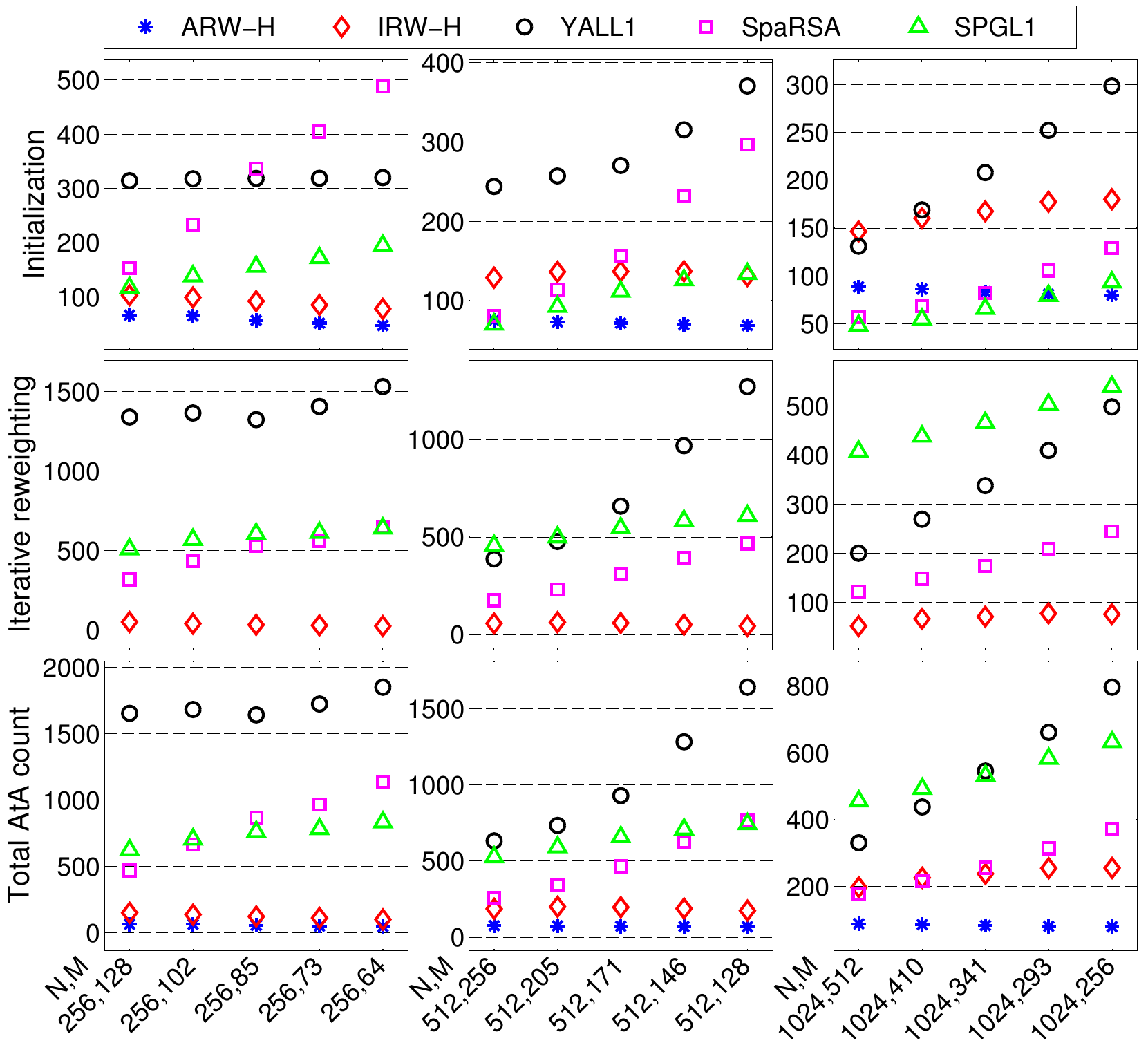}
    \caption{Comparison of the number of matrix-vector products for the recovery of HeaviSine signals.
    ARW-H solves adaptive-reweighted $\ell_1$ problem once, while other methods solve unweighted $\ell_1$ problem in their first iteration and perform five reweighted iterations afterwards.
    {\bf (First row)} Count for the first iteration only. {\bf (Second row)} Count for all the reweighting iterations (ARW-H does not appear because its count is zero). {\bf (Third row)} Count for all the iterations.}
    \label{fig:sType-HeaviSine_AtA}
\end{figure}

\else

\begin{figure*}
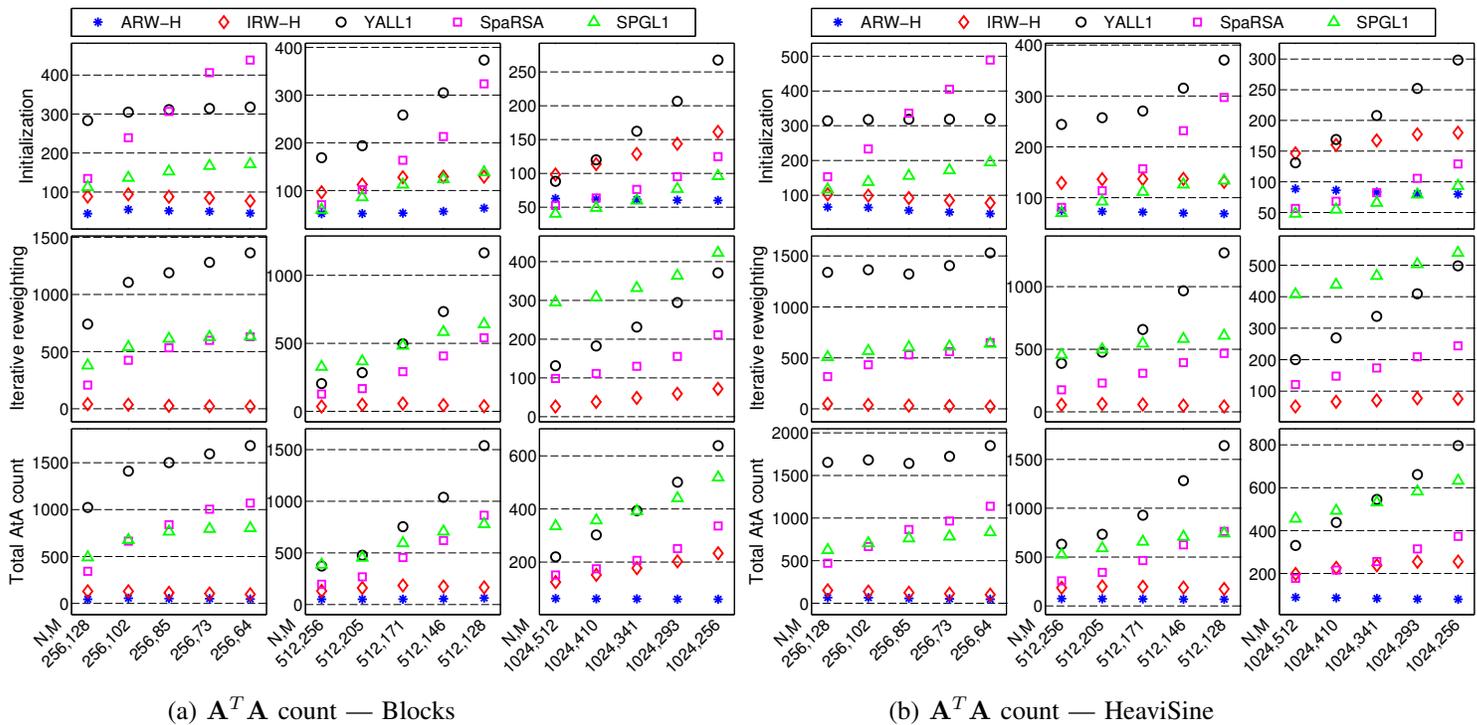

 \hspace{-35pt}
      \begin{subfigure}[t]{\figwidth}
        \centering
        \includegraphics[page=5,trim = 0mm 0mm 0mm 0mm, clip,  width=1.2\textwidth, keepaspectratio]{results_publish}
        \caption{$\mA^T\mA$ count --- Blocks}
    \label{fig:sType-blocks_AtA}
    \end{subfigure}
    \hspace{40pt}
    \begin{subfigure}[t]{\figwidth}
        \centering
        \includegraphics[page=14,trim = 0mm 0mm 0mm 0mm, clip, width=1.2\textwidth, keepaspectratio]{results_publish}
        \caption{$\mA^T\mA$ count --- HeaviSine}
        \label{fig:sType-HeaviSine_AtA}
    \end{subfigure}
    ~ 
    ~ 
    \caption{Comparison of the number of matrix-vector products for the recovery.
    ARW-H solves adaptive-reweighted $\ell_1$ problem once, while other methods solve unweighted $\ell_1$ problem in their first iteration and perform five reweighted iterations afterwards.
    {\bf (First row)} Count for the first iteration only. {\bf (Second row)} Count for all the reweighting iterations (ARW-H does not appear because its count is zero). {\bf (Third row)} Count for all the iterations.}
    \label{fig:AtA}
\end{figure*}

\fi

Comparison of MATLAB runtime for all the algorithms is presented in Fig.~\ref{fig:sType-blocks_CPU} (for Blocks signals) and Fig.~\ref{fig:sType-HeaviSine_CPU} (for HeaviSine signals).
The first row presents runtime that each algorithm utilized for computing the initial solution, the second row presents execution time for five reweighting iterations, and the third row presents total time consumed by each of the recovery algorithms. As we can see in the second row that, compared to YALL1, SPGL1, and SpaRSA, IRW-H consumed distinctly lesser time for updating solutions in iterative reweighting.
In the third row, we see small difference in the total runtime for IRW-H, SpaRSA, and YALL1, where IRW-H and SpaRSA display comparable performance.
Nevertheless, in all the experiments, the total runtime for ARW-H is the smallest among all the algorithms.

\iffullfigure
\begin{figure}[t]
    \centering
    \includegraphics[page=6,trim = 0mm 0mm 0mm 0mm, clip,  width=1\columnwidth]{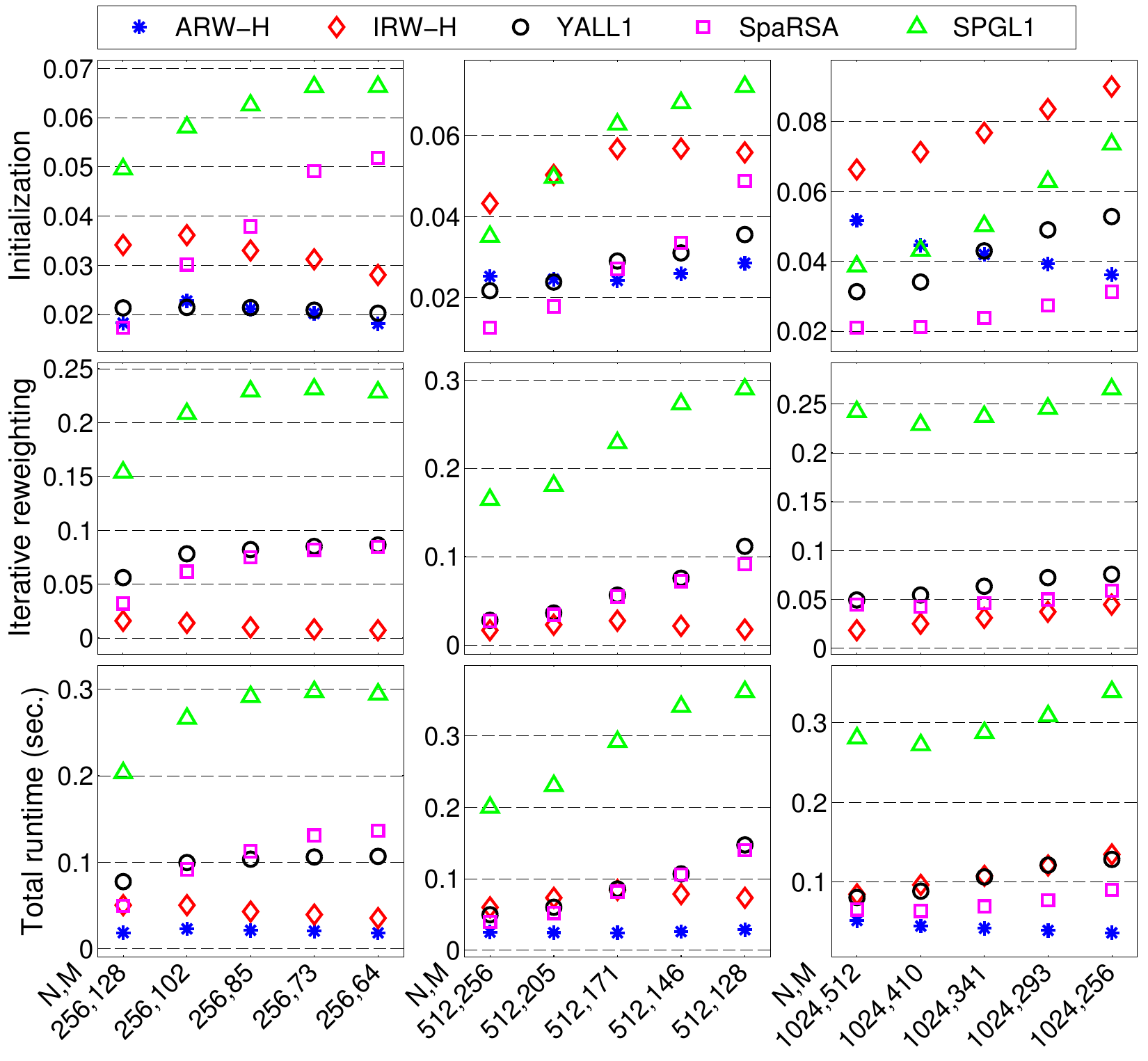}
    \caption{Comparison of MATLAB runtime for the recovery of Blocks signals.
    ARW-H solves adaptive-reweighted $\ell_1$ problem once, while other methods solve unweighted $\ell_1$ problem in their first iteration and perform five reweighted iterations afterwards.
    {\bf (First row)} Time for the first iteration. {\bf (Second row)} Time for all the reweighted iterations. {\bf (Third row)} Total runtime.}
    \label{fig:sType-blocks_CPU}
\end{figure}
\begin{figure}[t]
    \centering
    \includegraphics[page=15,trim = 0mm 0mm 0mm 0mm, clip, width=1\columnwidth, keepaspectratio]{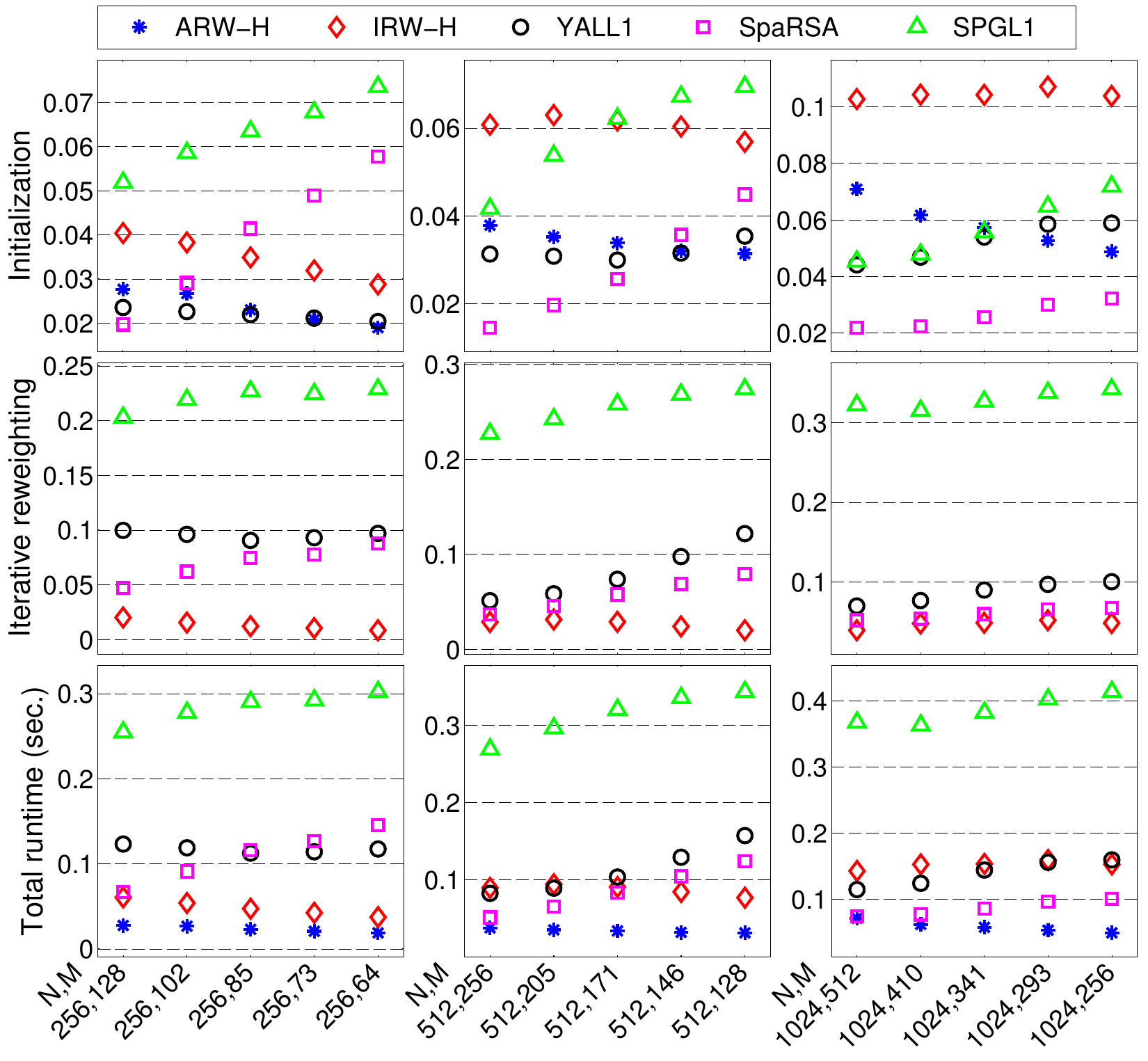}
    \caption{Comparison of MATLAB runtime for the recovery of HeaviSine signals. {\bf (First row)} Time for the first iteration. ARW-H solves adaptive-reweighted $\ell_1$ problem once, while other methods solve unweighted $\ell_1$ problem in their first iteration and perform five reweighted iterations afterwards.
    {\bf (Second row)} Time for all the reweighted iterations. {\bf (Third row)} Total runtime.}
    \label{fig:sType-HeaviSine_CPU}
\end{figure}

\else

\begin{figure*}
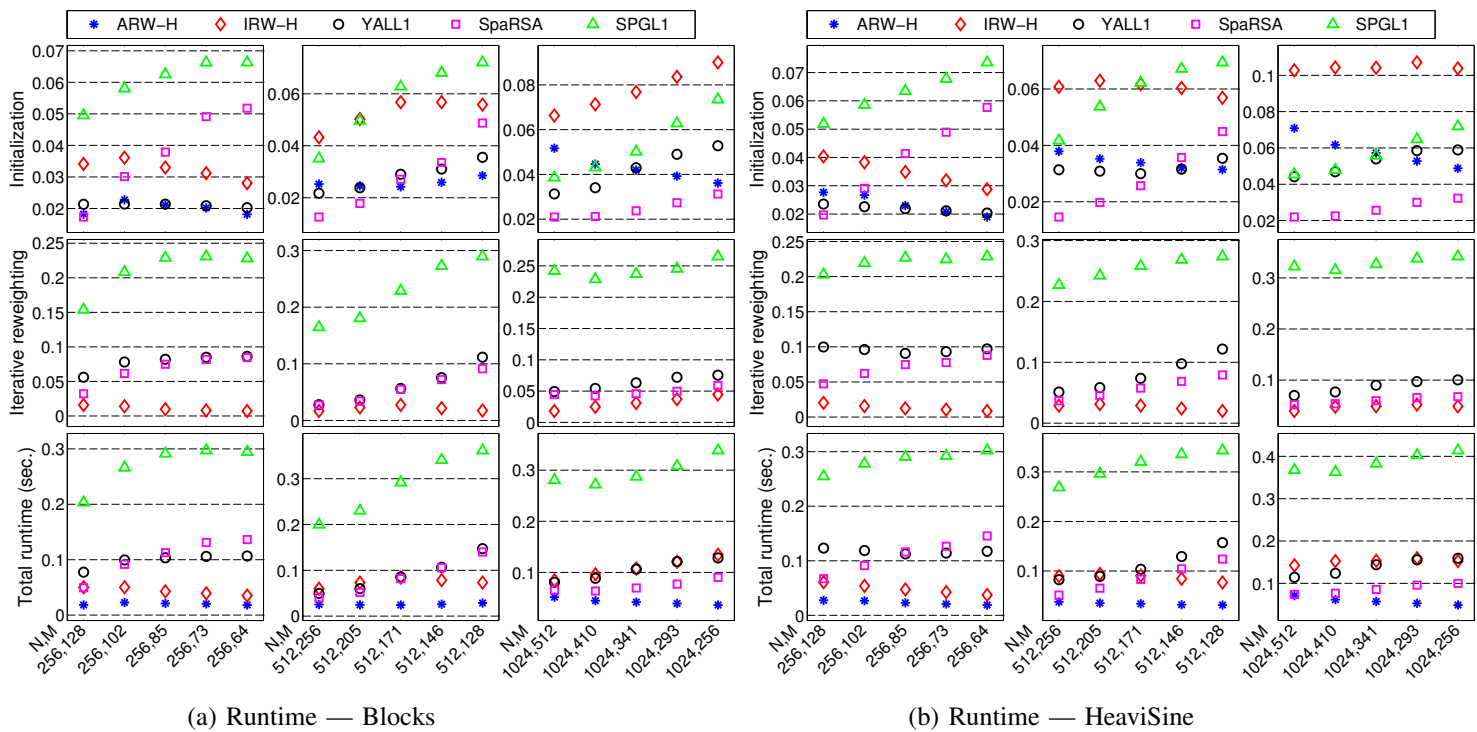

\hspace{-35pt}
      \begin{subfigure}[t]{\figwidth}
        \centering
        \includegraphics[page=6,trim = 0mm 0mm 0mm 0mm, clip,  width=1.2\textwidth, keepaspectratio]{results_publish}
        \caption{Runtime --- Blocks}
    \label{fig:sType-blocks_CPU}
    \end{subfigure}
    \hspace{40pt}
    \begin{subfigure}[t]{\figwidth}
        \centering
        \includegraphics[page=15,trim = 0mm 0mm 0mm 0mm, clip, width=1.2\textwidth, keepaspectratio]{results_publish}
        \caption{Runtime --- HeaviSine}
    \label{fig:sType-HeaviSine_CPU}
    \end{subfigure}
    ~ 
    ~ 
    \caption{Comparison of MATLAB runtime for the recovery. {\bf (First row)} Time for the first iteration. ARW-H solves adaptive-reweighted $\ell_1$ problem once, while other methods solve unweighted $\ell_1$ problem in their first iteration and perform five reweighted iterations afterwards.
    {\bf (Second row)} Time for all the reweighted iterations. {\bf (Third row)} Total runtime.}
    \label{fig:CPU}
\end{figure*}

\fi

A brief summary of the results for our experiments is as follows. We observed that the adaptive reweighting scheme (ARW-H) recovered signals with better SERs compared to the iterative reweighting schemes (IRW-H, SpaRSA, YALL1, and SPGL1), and it does so by solving a single homotopy problem at the expense of a small amount of computational cost and time. Among the iterative reweighting schemes, IRW-H quickly updated the solutions during iterative reweighting at the expense of marginal computational cost and time, which are distinctly smaller than the respective costs and times for SpaRSA, YALL1, and SPGL1; although SpaRSA and YALL1 with warm-start provided competitive results for longer signals.

\section{Discussion}
We presented two homotopy algorithms that can efficiently solve reweighted $\ell_1$ problems. In Sec.~\ref{sec:rwt}, we presented an algorithm for updating the solution of \eqref{eq:wtBPDN} as the $\vw_i$ change. We demonstrated with experiments that, in reweighting iterations, our proposed algorithm quickly updates the solution at a small computational expense.
In Sec.~\ref{sec:adp}, we presented a homotopy algorithm that adaptively selects weights inside a single homotopy program. We demonstrated with experiments that our proposed adaptive reweighting method outperforms iterative reweighting methods in terms of the reconstruction quality and the computational cost.

We would like to mention that an adaptive reweighting scheme, similar to the one we presented for homotopy, can also be embedded inside iterative shrinkage-thresholding algorithms \cite{Daubechies_2004_ItrativeThresholding,Wright-2009-sparsa,bioucas-2007-TwIST,beck-2009-FISTA,hale2008fpc}. A recent paper \cite{Mansour-2012-WSPGL1} has also employed a similar principle of reweighting inside SPGL1.
Standard iterative shrinkage-thresholding algorithms solve the program in \eqref{eq:wtBPDN} by solving the following shrinkage problem at every inner iteration:
\begin{equation}\label{eq:ISTA-soft}
\underset{\vx}{\text{minimize}}\; \frac{L}{2}\|\vx - \vu\|_2^2 + \sum \vw_i |\vx_i|,
\end{equation}
where $\vu = \vx^{k-1} - \frac{1}{L}\mA^T(\mA\vx^{k-1}-\vy)$ denotes a vector that is generated using a solution $\vx^{k-1}$ from a previous iteration and $L$ determines the stepsize. The solution of \eqref{eq:ISTA-soft} involves a simple soft-thresholding of the entries in $\vu$ with respect to the $\vw_i/L$ (i.e., $\vx_i = \text{soft}(\vu_i, \vw_i/L)$, where $\text{soft}(u,\alpha) \equiv \sign{u}\max\{|u|-\alpha,0\}$ defines the soft-thresholding/shrinkage function).
To embed adaptive reweighting inside such shrinkage algorithms, instead of using a fixed set of $\vw_i$ for soft-thresholding in \eqref{eq:ISTA-soft} at every iteration, we can adaptively select the $\vw_i$ according to the changes in the solution.

\newif\ifLargeM
\newif\ifdaubOrth

\LargeMtrue 

\daubOrthtrue 
\daubOrthfalse 

To examine our suggestion, we added an adaptive reweighting scheme in the source code of SpaRSA. SpaRSA offers a feature for adaptive continuation in which it starts the shrinkage parameter $\tau$ with a large value and decreases it towards the desired value after every few iterations.
We added an extra line in the code that uses the available solution $\hat \vx$ to update each $\vw_i$ as $\min\left(\tau,\tau/\beta|\hat\vx_i|\right)$ whenever $\tau$ changes.
We gauged the performance of this modified method by using it for the recovery of gray-scale images from compressive measurements. The problem setup following the model in \eqref{eq:y=Ax+e} is as follows. We generated a sparse signal $\bar \vx$ of length $N$ by applying a
\ifdaubOrth Daubechies~4 orthogonal wavelet transform~\cite{Mallat_book_WaveletTour99}
\else Daubechies~9/7 biorthogonal wavelet transform~\cite{CDF-1992-Biorthogonal,Mallat_book_WaveletTour99} with odd-symmetric extension
\fi
on an image, selected $\mA$ as a subsampled noiselet transform~\cite{Coifman_2001_Noiselets}, and added Gaussian noise $\ve$ in the measurements by selecting each entry in $\ve$ as i.i.d. $\mathcal{N}(0,\sigma^2)$.

In Fig.~\ref{fig:largescale} we present results averaged over 10 experiments that we performed for the recovery of three $256\times 256$ images from \ifLargeM $M = 30,000$ \else $M = 25,000$ \fi noiselet measurements in the presence of noise at $40$\,dB SNR using SpaRSA in three different ways.
The first column presents the original $256\times256$ images.
The second column presents small portions of the images that we reconstructed by solving the standard $\ell_1$ problem in \eqref{eq:BPDN} using $\tau = \sigma \sqrt{\log N}$. The peak signal-to-noise ratio (PSNR) for the entire reconstructed image is presented in each caption along with the total count for the number of applications of $\mA^T\mA$ (in parentheses) averaged over 10 experiments.
The third column presents portions of the reconstructed images after three reweighting iterations using the warm-start and weight selection procedure employed in the experiments in Sec.~\ref{sec:exp}.
The last column presents portions of the reconstructed images by solving the modified version of SpaRSA in which we initialized the SpaRSA code by setting all the weights to a same value, and after every continuation iteration we modified the values of weights according to the available solution. We observe that SpaRSA with this adaptive reweighting modification yields better performance in terms of PSNR over the other two methods, while its computational cost is significantly smaller than the cost of iterative reweighting (in column 2).

We have presented these results as a proof-of-concept and we anticipate that adding a simple adaptive reweighting scheme within existing iterative shrinkage algorithms can potentially enhance their performance in many scenarios without any additional cost; however, a detailed study in this regard is beyond the scope of this paper.


\renewcommand{\figwidth}{0.24\textwidth}

\ifdaubOrth
\begin{figure*}
\centering
    \def\llx{0.51}
    \def\lly{0.51}
    \def\urx{0.95}
    \def\ury{0.95}
    \begin{subfigure}[t]{\figwidth}
        \centering
        \begin{tikzpicture}
        \node[anchor=south west,inner sep=0] (image) at (0,0) {\includegraphics[width=\textwidth]{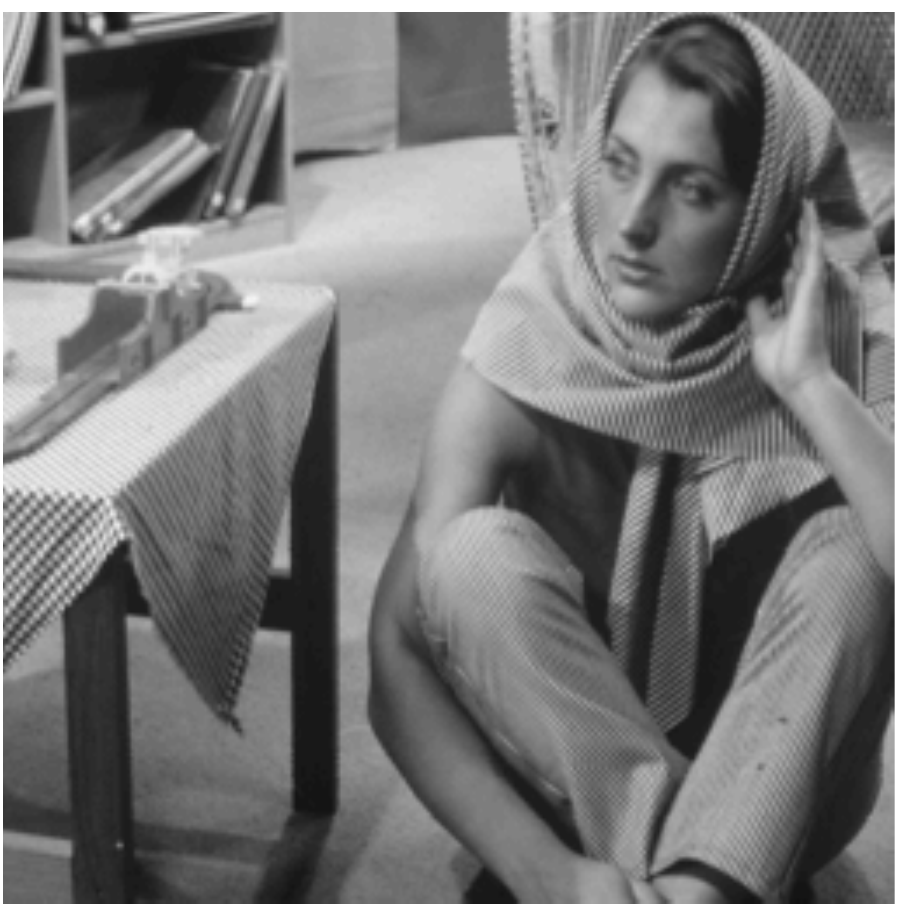}};
        \begin{scope}[x={(image.south east)},y={(image.north west)}]
        \draw[orange, line width = 1.5pt] (\llx,\lly) rectangle (\urx,\ury);
        \end{scope}
        \end{tikzpicture}
        \captionsetup{font=footnotesize}
        \caption{Barbara}
        \label{fig:barbara1}
    \end{subfigure}
    \begin{subfigure}[t]{\figwidth}
        \centering
        \ifLargeM
        \includegraphics[width=\textwidth, rviewport= {\llx} {\lly} {\urx} {\ury}, clip]{barbara_N65536_M30000_SNR40_adp0_rwt1_daub4}
        \captionsetup{font=footnotesize}
        \caption{PSNR: $27.87$\,dB --- ($64$) }
        \else
        \includegraphics[width=\textwidth, rviewport= {\llx} {\lly} {\urx} {\ury}, clip]{barbara_N65536_M25000_SNR40_adp0_rwt1_daub4} \captionsetup{font=footnotesize}
        \caption{PSNR: $26.35$\,dB --- ($74$)}
        \fi
        \label{fig:barbara2}
    \end{subfigure}
    \begin{subfigure}[t]{0.24\textwidth}
        \centering
        \ifLargeM
        \includegraphics[width=\textwidth, rviewport= {\llx} {\lly} {\urx} {\ury}, clip]{barbara_N65536_M30000_SNR40_adp0_rwt4_daub4}
        \captionsetup{font=footnotesize}
        \caption{PSNR: $27.7$\,dB --- ($259$)}
        \else
        \includegraphics[width=\textwidth, rviewport= {\llx} {\lly} {\urx} {\ury}, clip]{barbara_N65536_M25000_SNR40_adp0_rwt4_daub4} \captionsetup{font=footnotesize}
        \caption{PSNR: $26.04$\,dB --- ($300$) }
        \fi
        \label{fig:barbara3}
    \end{subfigure}
    \begin{subfigure}[t]{\figwidth}
        \centering
        \ifLargeM
        \includegraphics[width=\textwidth, rviewport= {\llx} {\lly} {\urx} {\ury}, clip]{barbara_N65536_M30000_SNR40_adp1_rwt1_daub4}
        \captionsetup{font=footnotesize}
        \caption{PSNR: $28.29$\,dB --- ($99$)}
        \else
        \includegraphics[width=\textwidth, rviewport= {\llx} {\lly} {\urx} {\ury}, clip]{barbara_N65536_M25000_SNR40_adp1_rwt1_daub4}
        \captionsetup{font=footnotesize}
        \caption{PSNR: $26.58$\,dB --- ($105$)}
        \fi
        \label{fig:barbara4}
    \end{subfigure}

    \vspace{1pt}

    \def\llx{0.46}
    \def\lly{0.36}
    \def\urx{0.95}
    \def\ury{0.85}
    \begin{subfigure}[t]{\figwidth}
        \centering
        \begin{tikzpicture}
        \node[anchor=south west,inner sep=0] (image) at (0,0) {\includegraphics[width=\textwidth]{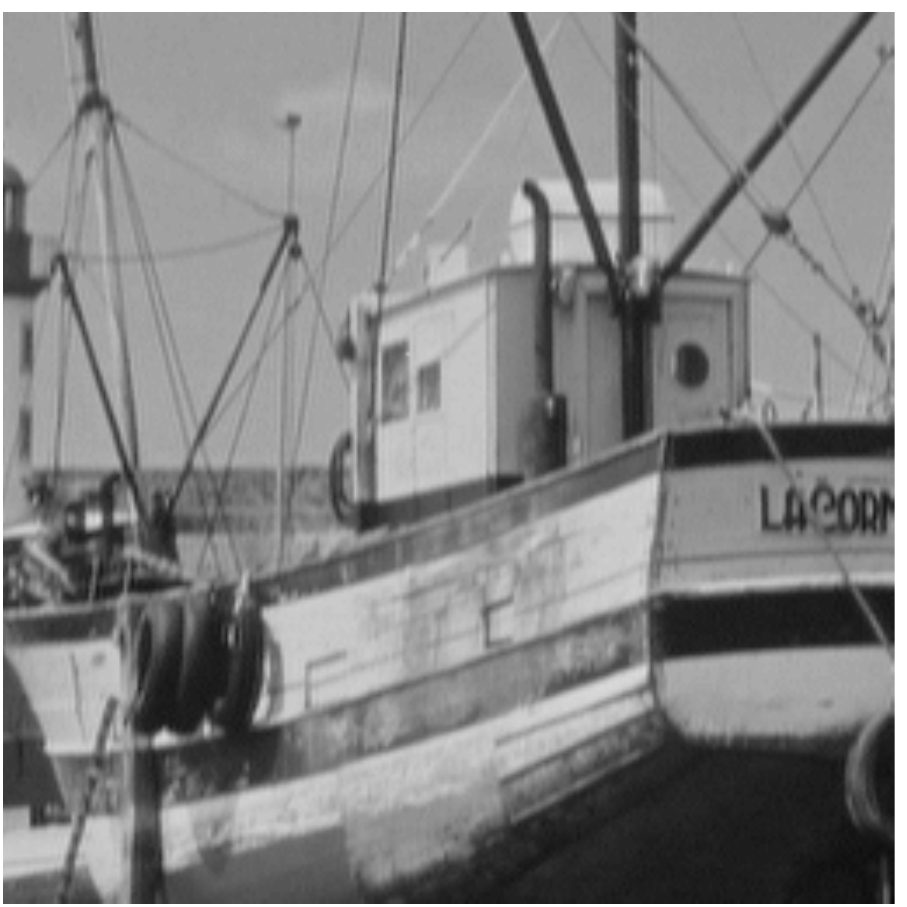}};
        \begin{scope}[x={(image.south east)},y={(image.north west)}]
        \draw[orange, line width = 1.5pt] (\llx,\lly) rectangle (\urx,\ury);
        \end{scope}
        \end{tikzpicture}
        \captionsetup{font=footnotesize}
        \caption{Boats}
        \label{fig:boats1}
    \end{subfigure}
    \begin{subfigure}[t]{\figwidth}
        \centering
        \ifLargeM
        \includegraphics[width=\textwidth, rviewport= {\llx} {\lly} {\urx} {\ury}, clip]{boats_N65536_M30000_SNR40_adp0_rwt1_daub4}
        \captionsetup{font=footnotesize}
        \caption{PSNR: $28.05$\,dB --- ($56$)}
        \else
        \includegraphics[width=\textwidth, rviewport= {\llx} {\lly} {\urx} {\ury}, clip]{boats_N65536_M25000_SNR40_adp0_rwt1_daub4}
        \captionsetup{font=footnotesize}
        \caption{PSNR: $26.4$\,dB --- ($64$)}
        \fi
        \label{fig:boats2}
    \end{subfigure}
    \begin{subfigure}[t]{\figwidth}
        \centering
        \ifLargeM
        \includegraphics[width=\textwidth, rviewport= {\llx} {\lly} {\urx} {\ury}, clip]{boats_N65536_M30000_SNR40_adp0_rwt4_daub4}
        \captionsetup{font=footnotesize}
        \caption{PSNR: $28.25$\,dB --- ($209$)}
        \else
        \includegraphics[width=\textwidth, rviewport= {\llx} {\lly} {\urx} {\ury}, clip]{boats_N65536_M25000_SNR40_adp0_rwt4_daub4}
        \captionsetup{font=footnotesize}
        \caption{PSNR: $26.34$\,dB --- ($252$) }
        \fi
        \label{fig:boats3}
    \end{subfigure}
    \begin{subfigure}[t]{\figwidth}
        \centering
        \ifLargeM
        \includegraphics[width=\textwidth, rviewport= {\llx} {\lly} {\urx} {\ury}, clip]{boats_N65536_M30000_SNR40_adp1_rwt1_daub4}
        \captionsetup{font=footnotesize}
        \caption{PSNR: $28.9$\,dB --- ($85$)}
        \else
        \includegraphics[width=\textwidth, rviewport= {\llx} {\lly} {\urx} {\ury}, clip]{boats_N65536_M25000_SNR40_adp1_rwt1_daub4}
        \captionsetup{font=footnotesize}
        \caption{PSNR: $27.01$\,dB --- ($91$)}
        \fi
        \label{fig:boats4}
    \end{subfigure}

    \vspace{1pt}

    \def\llx{0.31}
    \def\lly{0.36}
    \def\urx{0.80}
    \def\ury{0.85}
    \begin{subfigure}[t]{\figwidth}
        \centering
        \begin{tikzpicture}
        \node[anchor=south west,inner sep=0] (image) at (0,0) {\includegraphics[width=\textwidth]{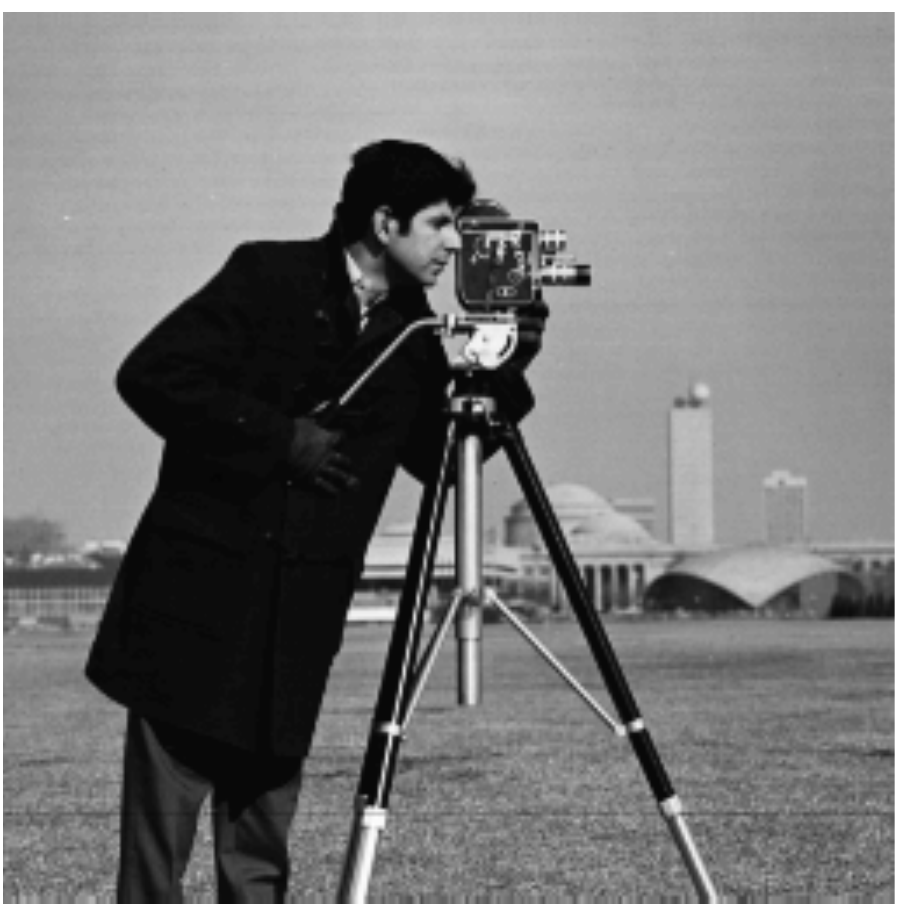}};
        \begin{scope}[x={(image.south east)},y={(image.north west)}]
        \draw[orange, line width = 1.5pt] (\llx,\lly) rectangle (\urx,\ury);
        \end{scope}
        \end{tikzpicture}
        \captionsetup{font=footnotesize}
        \caption{Cameraman}
        \label{fig:cameraman1}
    \end{subfigure}
    \begin{subfigure}[t]{\figwidth}
        \centering
        \ifLargeM
        \includegraphics[width=\textwidth, rviewport= {\llx} {\lly} {\urx} {\ury}, clip]{cameraman_N65536_M30000_SNR40_adp0_rwt1_daub4}
        \captionsetup{font=footnotesize}
        \caption{PSNR: $29.08$\,dB --- ($59$)}
        \else
        \includegraphics[width=\textwidth, rviewport= {\llx} {\lly} {\urx} {\ury}, clip]{cameraman_N65536_M25000_SNR40_adp0_rwt1_daub4}
        \captionsetup{font=footnotesize}
        \caption{PSNR: $27.28$\,dB --- ($72$) }
        \fi
        \label{fig:cameraman2}
    \end{subfigure}
    \begin{subfigure}[t]{\figwidth}
        \centering
        \ifLargeM
        \includegraphics[width=\textwidth, rviewport= {\llx} {\lly} {\urx} {\ury}, clip]{cameraman_N65536_M30000_SNR40_adp0_rwt4_daub4}
        \captionsetup{font=footnotesize}
        \caption{PSNR: $29.27$\,dB --- ($221$)}
        \else
        \includegraphics[width=\textwidth, rviewport= {\llx} {\lly} {\urx} {\ury}, clip]{cameraman_N65536_M25000_SNR40_adp0_rwt4_daub4}
        \captionsetup{font=footnotesize}
        \caption{PSNR: $27.14$\,dB --- ($273$) }
        \fi
        \label{fig:cameraman3}
    \end{subfigure}
    \begin{subfigure}[t]{\figwidth}
        \centering
        \ifLargeM
        \includegraphics[width=\textwidth, rviewport= {\llx} {\lly} {\urx} {\ury}, clip]{cameraman_N65536_M30000_SNR40_adp1_rwt1_daub4}
        \captionsetup{font=footnotesize}
        \caption{PSNR: $29.98$\,dB --- ($86$)}
        \else
        \includegraphics[width=\textwidth, rviewport= {\llx} {\lly} {\urx} {\ury}, clip]{cameraman_N65536_M25000_SNR40_adp1_rwt1_daub4}
        \captionsetup{font=footnotesize}
        \caption{PSNR: $27.90$\,dB --- ($96$) }
        \fi
        \label{fig:cameraman4}
    \end{subfigure}

    ~ 
    ~ 

    \caption{Results for the recovery of $256\times256$ images from \ifLargeM $M=30,000$ \else $M = 25,000$ \fi noiselet measurements in the presence of Gaussian noise at $40${}dB SNR, using Daubechies~4 orthogonal wavelet transform as the sparse representation. \textbf{(Column~1)} Original images. \textbf{ (Column~2)} Portions of the images (inside the orange box) reconstructed by solving \eqref{eq:BPDN} using SpaRSA. \textbf{(Column~3)} Reconstruction after three reweighting iterations. \textbf{(Column~4)} Adaptive reweighting by updating the $\vw_i$ after every continuation step in SpaRSA.  The caption under each subimage shows the PSNR over the entire reconstructed image and a count for the number of applications of $\mA^T\mA$ (in parentheses)  averaged over 10 experiments.}
    \label{fig:largescale}
\end{figure*}

\else

\begin{figure*}
\centering
    \def\llx{0.51}
    \def\lly{0.51}
    \def\urx{0.98}
    \def\ury{0.98}
    \begin{subfigure}[t]{\figwidth}
        \centering
        \begin{tikzpicture}
        \node[anchor=south west,inner sep=0] (image) at (0,0) {\includegraphics[width=1\textwidth]{barbara}};
        \begin{scope}[x={(image.south east)},y={(image.north west)}]
        \draw[orange, line width = 1.5pt] (\llx,\lly) rectangle (\urx,\ury);
        \end{scope}
        \end{tikzpicture}
        \captionsetup{font=footnotesize}
        \caption{Barbara}
        \label{fig:barbara1}
    \end{subfigure}
    \begin{subfigure}[t]{\figwidth}
        \centering
        \ifLargeM
        \includegraphics[width=\textwidth, rviewport= {\llx} {\lly} {\urx} {\ury}, clip]{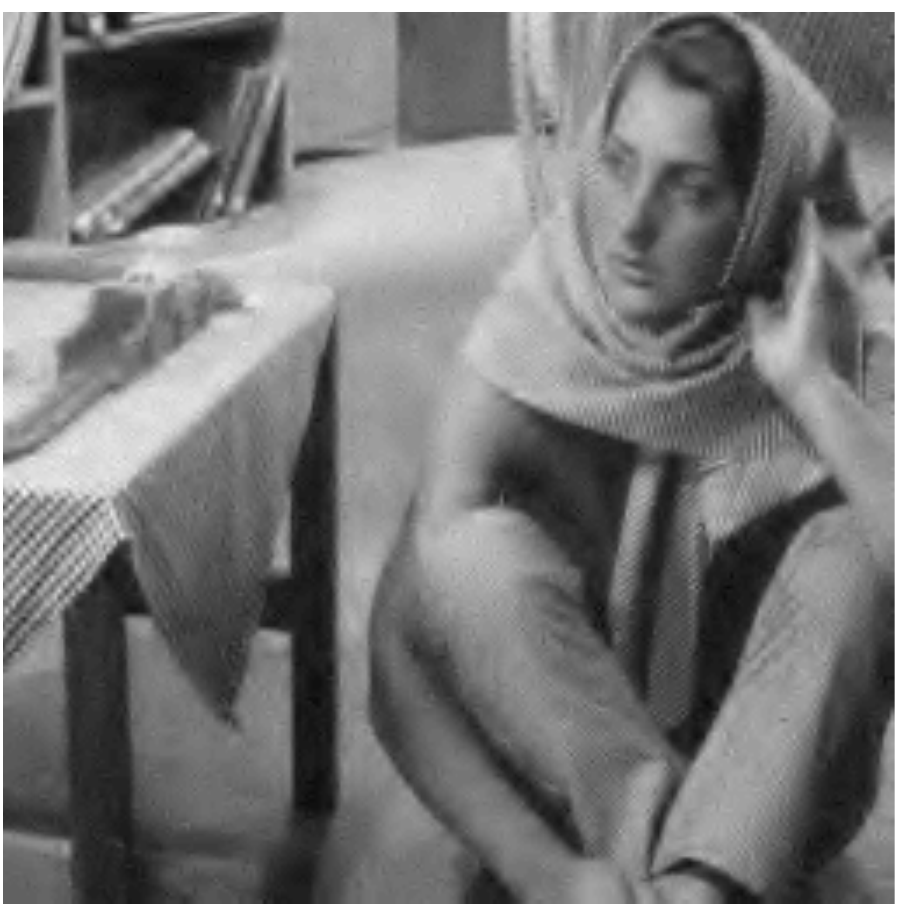}
        \captionsetup{font=footnotesize}
        \caption{PSNR: $29.05$\,dB --- ($55$) }
        \else
        \includegraphics[width=\textwidth, rviewport= {\llx} {\lly} {\urx} {\ury}, clip]{barbara_N65536_M25000_SNR40_adp0_rwt1_daub79} \captionsetup{font=footnotesize}
        \caption{PSNR: $27.62$\,dB --- ($62$) }
        \fi
        \label{fig:barbara2}
    \end{subfigure}
    \begin{subfigure}[t]{\figwidth}
        \centering
        \ifLargeM
        \includegraphics[width=\textwidth, rviewport= {\llx} {\lly} {\urx} {\ury}, clip]{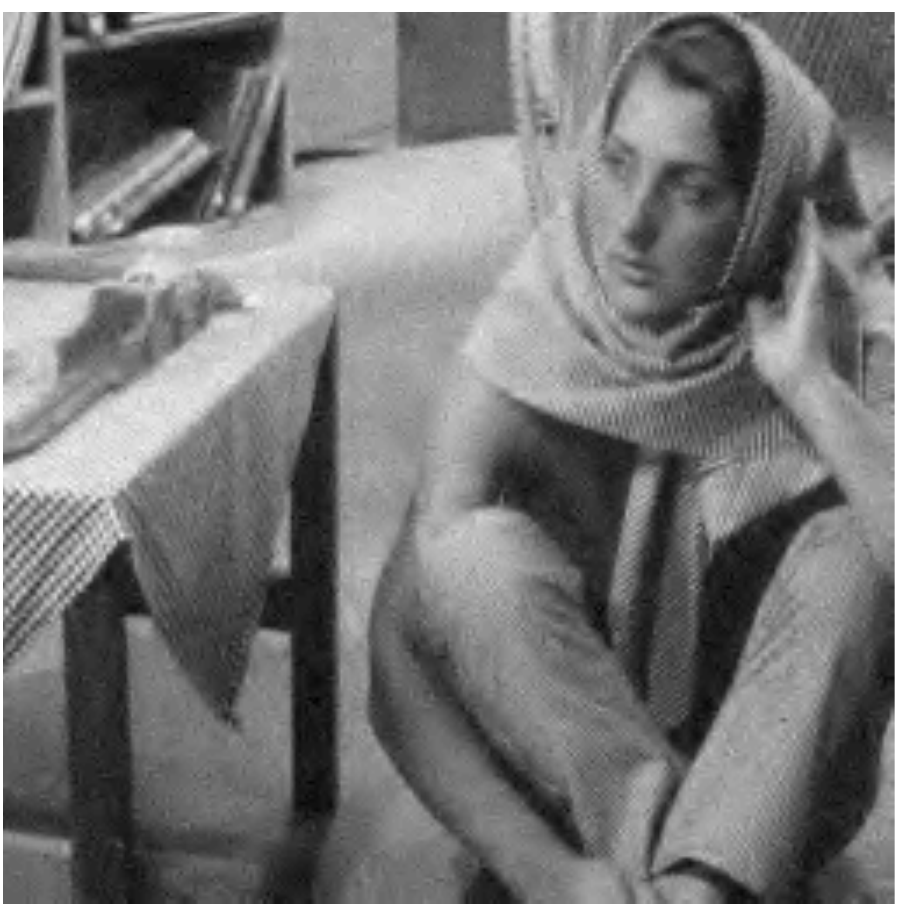}
        \captionsetup{font=footnotesize}
        \caption{PSNR: $29.18$\,dB --- ($209$) }
        \else
        \includegraphics[width=\textwidth, rviewport= {\llx} {\lly} {\urx} {\ury}, clip]{barbara_N65536_M25000_SNR40_adp0_rwt4_daub79} \captionsetup{font=footnotesize}
        \caption{PSNR: $27.50$\,dB --- ($250$) }
        \fi
        \label{fig:barbara3}
    \end{subfigure}
    \begin{subfigure}[t]{\figwidth}
        \centering
        \ifLargeM
        \includegraphics[width=\textwidth, rviewport= {\llx} {\lly} {\urx} {\ury}, clip]{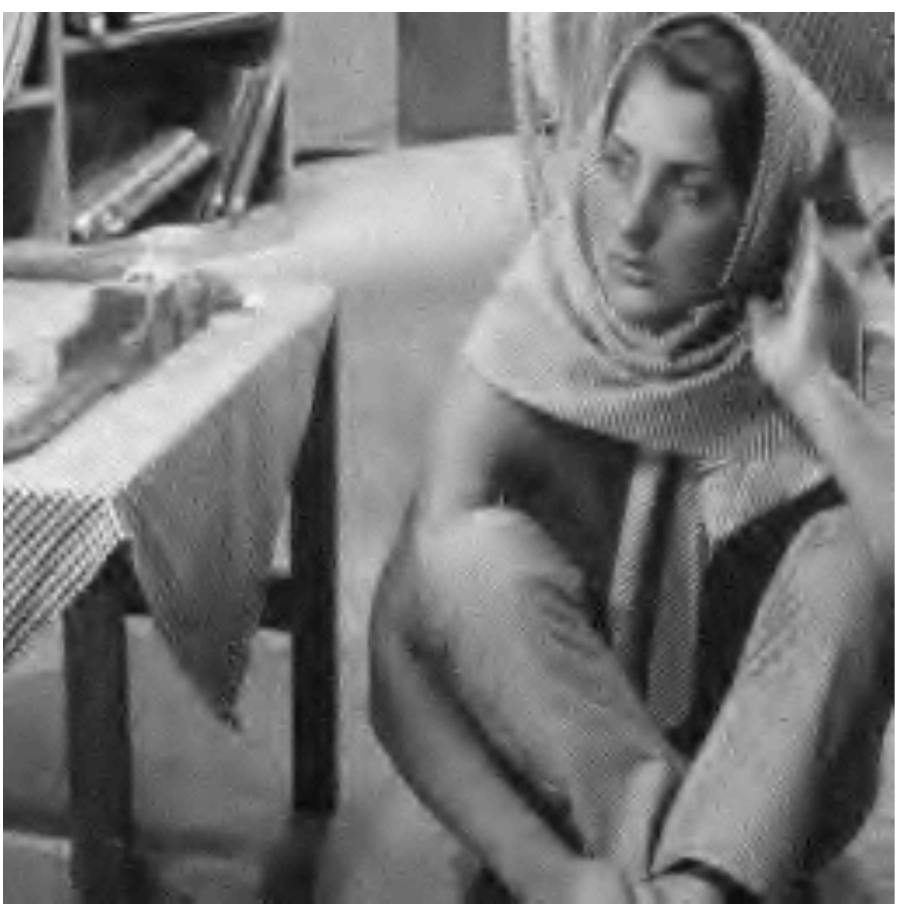}
        \captionsetup{font=footnotesize}
        \caption{PSNR: $29.68$\,dB --- ($85$) }
        \else
        \includegraphics[width=\textwidth, rviewport= {\llx} {\lly} {\urx} {\ury}, clip]{barbara_N65536_M25000_SNR40_adp1_rwt1_daub79}
        \captionsetup{font=footnotesize}
        \caption{PSNR: $28.11$\,dB --- ($96$) }
        \fi
        \label{fig:barbara4}
    \end{subfigure}

    \vspace{1pt}

    \def\llx{0.41}
    \def\lly{0.31}
    \def\urx{0.95}
    \def\ury{0.85}
    \begin{subfigure}[t]{\figwidth}
        \centering
        \begin{tikzpicture}
        \node[anchor=south west,inner sep=0] (image) at (0,0) {\includegraphics[width=\textwidth]{boats}};
        \begin{scope}[x={(image.south east)},y={(image.north west)}]
        \draw[orange, line width = 1.5pt] (\llx,\lly) rectangle (\urx,\ury);
        \end{scope}
        \end{tikzpicture}
        \captionsetup{font=footnotesize}
        \caption{Boats}
        \label{fig:boats1}
    \end{subfigure}
    \begin{subfigure}[t]{\figwidth}
        \centering
        \ifLargeM
        \includegraphics[width=\textwidth, rviewport= {\llx} {\lly} {\urx} {\ury}, clip]{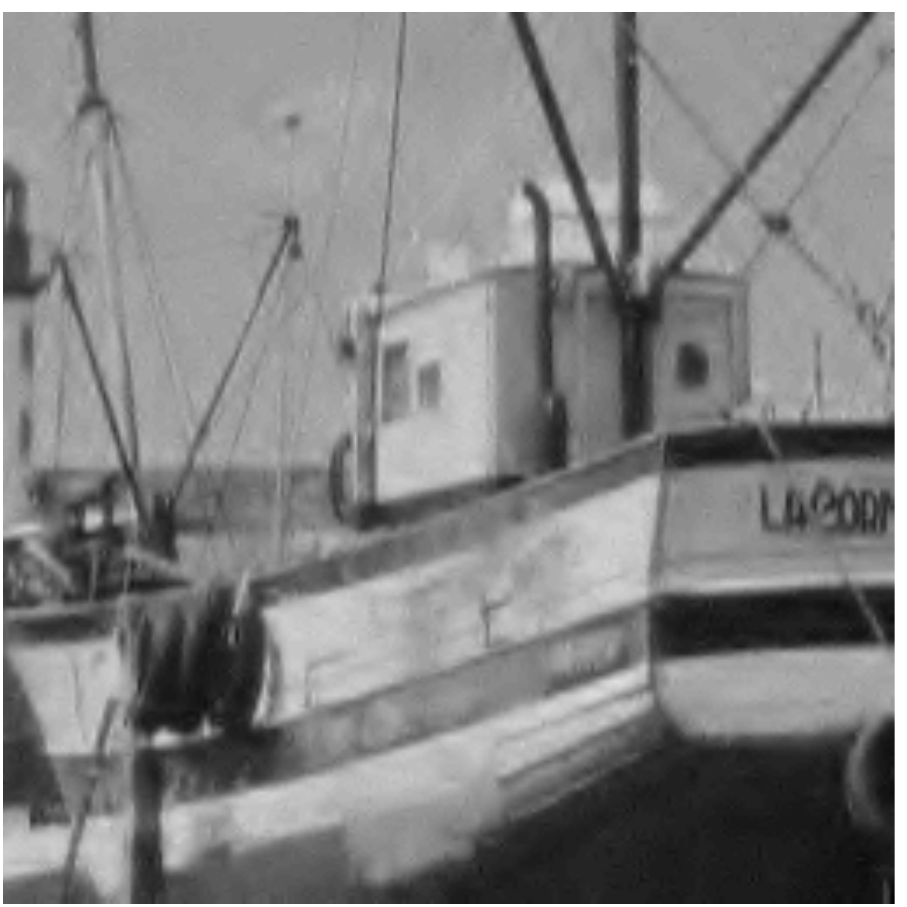}
        \captionsetup{font=footnotesize}
        \caption{PSNR: $29.23$\,dB --- ($49$) }
        \else
        \includegraphics[width=\textwidth, rviewport= {\llx} {\lly} {\urx} {\ury}, clip]{boats_N65536_M25000_SNR40_adp0_rwt1_daub79}
        \captionsetup{font=footnotesize}
        \caption{PSNR: $27.70$\,dB --- ($58$) }
        \fi
        \label{fig:boats2}
    \end{subfigure}
    \begin{subfigure}[t]{\figwidth}
        \centering
        \ifLargeM
        \includegraphics[width=\textwidth, rviewport= {\llx} {\lly} {\urx} {\ury}, clip]{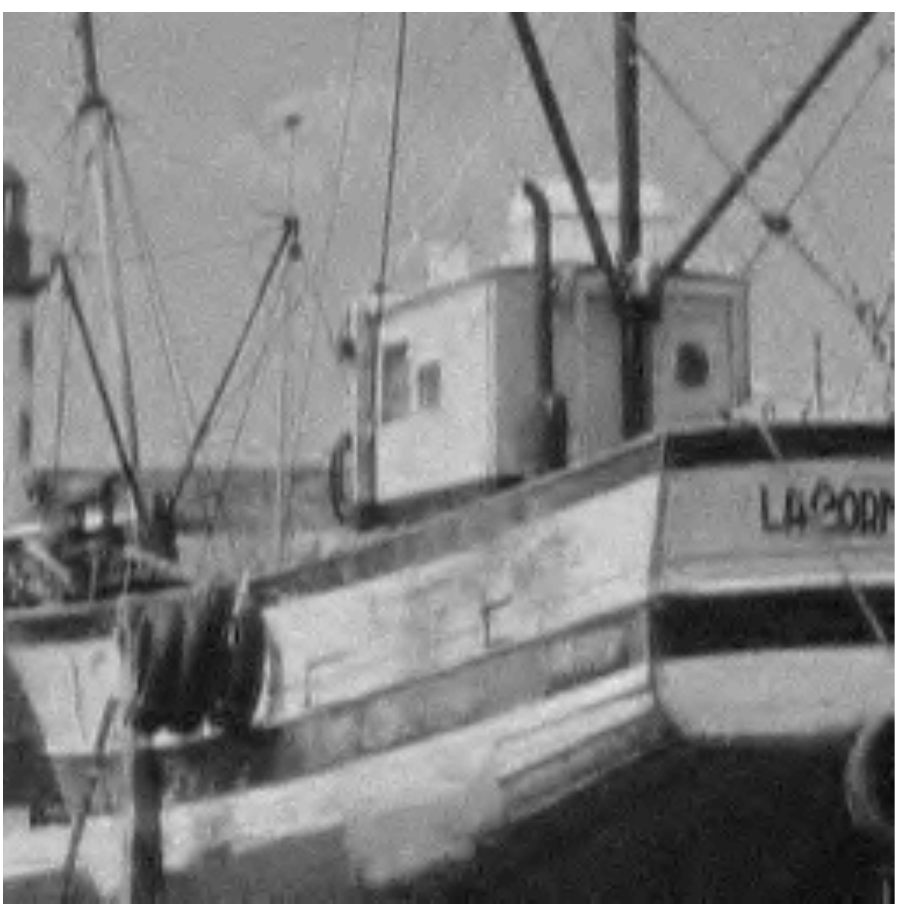}
        \captionsetup{font=footnotesize}
        \caption{PSNR: $30.00$\,dB --- ($173$) }
        \else
        \includegraphics[width=\textwidth, rviewport= {\llx} {\lly} {\urx} {\ury}, clip]{boats_N65536_M25000_SNR40_adp0_rwt4_daub79}
        \captionsetup{font=footnotesize}
        \caption{PSNR: $28.06$\,dB --- ($211$) }
        \fi
        \label{fig:boats3}
    \end{subfigure}
    \begin{subfigure}[t]{\figwidth}
        \centering
        \ifLargeM
        \includegraphics[width=\textwidth, rviewport= {\llx} {\lly} {\urx} {\ury}, clip]{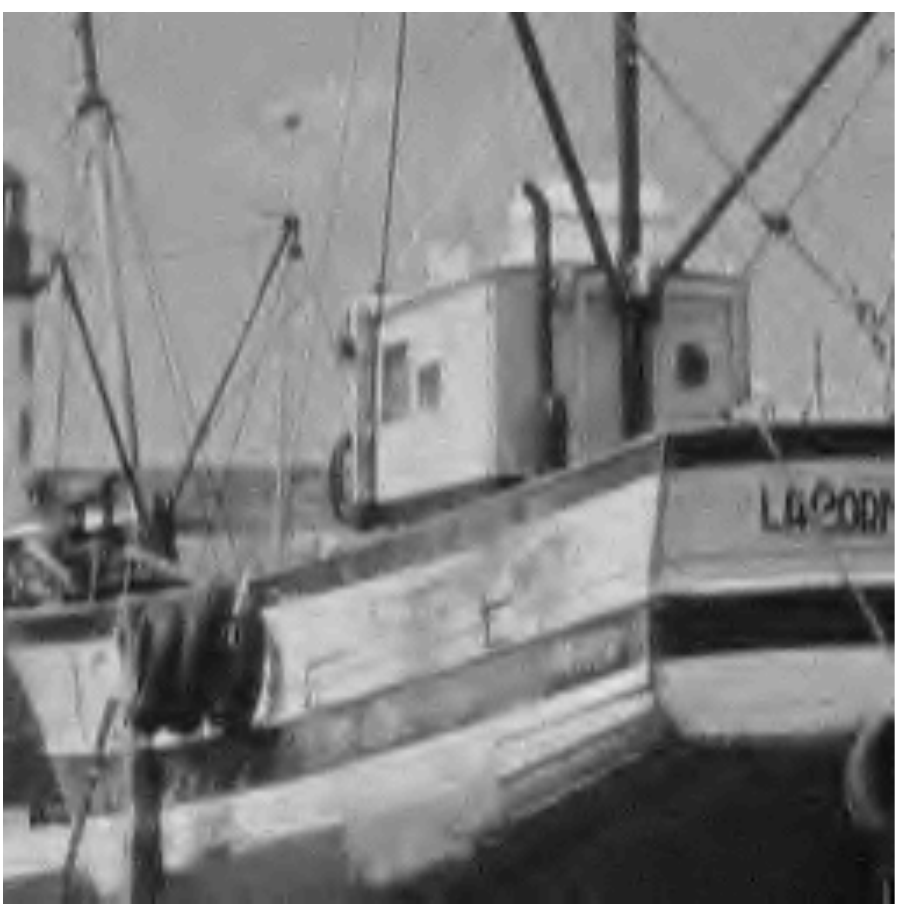}
        \captionsetup{font=footnotesize}
        \caption{PSNR: $30.28$\,dB --- ($72$) }
        \else
        \includegraphics[width=\textwidth, rviewport= {\llx} {\lly} {\urx} {\ury}, clip]{boats_N65536_M25000_SNR40_adp1_rwt1_daub79}
        \captionsetup{font=footnotesize}
        \caption{PSNR: $28.61$\,dB --- ($81$)}
        \fi
        \label{fig:boats4}
    \end{subfigure}

    \vspace{1pt}

    \def\llx{0.31}
    \def\lly{0.36}
    \def\urx{0.80}
    \def\ury{0.85}
    \begin{subfigure}[t]{\figwidth}
        \centering
        \begin{tikzpicture}
        \node[anchor=south west,inner sep=0] (image) at (0,0) {\includegraphics[trim= 0mm  0mm 0mm 0mm, clip, width=\textwidth]{cameraman}};
        \begin{scope}[x={(image.south east)},y={(image.north west)}]
        \draw[orange, line width = 1.5pt] (\llx,\lly) rectangle (\urx,\ury);
        \end{scope}
        \end{tikzpicture}
        \captionsetup{font=footnotesize}
        \caption{Cameraman}
        \label{fig:cameraman1}
    \end{subfigure}
    \begin{subfigure}[t]{\figwidth}
        \centering
        \ifLargeM
        \includegraphics[width=\textwidth, rviewport= {\llx} {\lly} {\urx} {\ury}, clip]{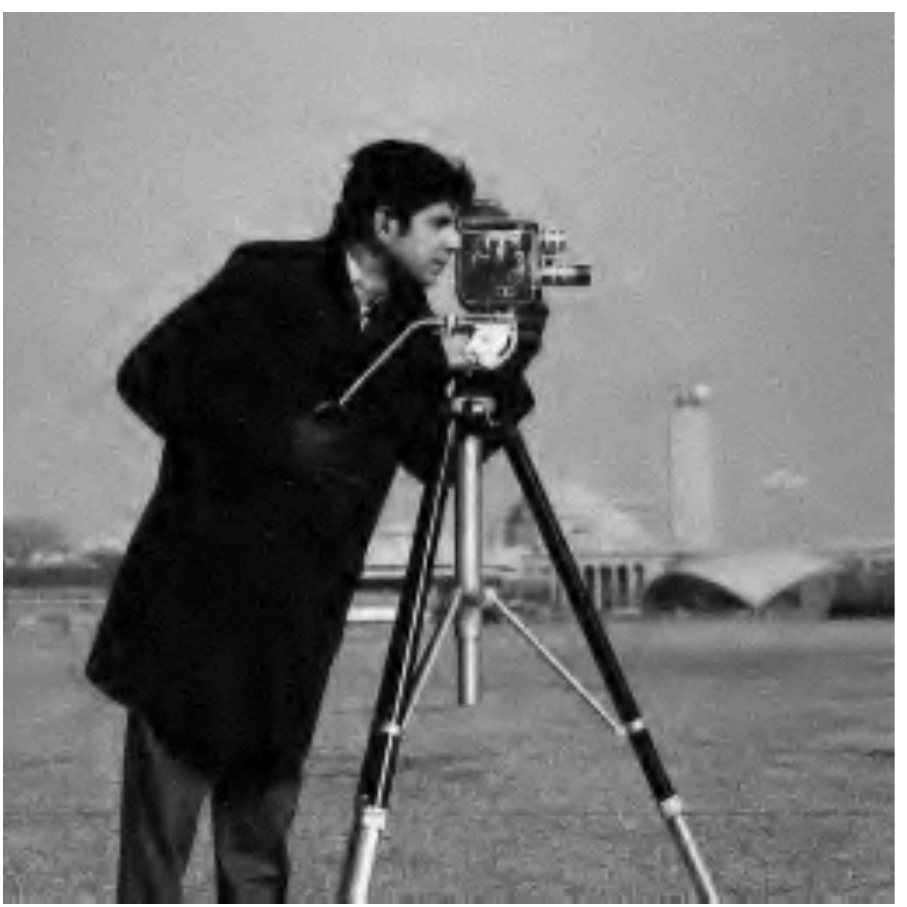}
        \captionsetup{font=footnotesize}
        \caption{PSNR: $29.82$\,dB --- ($52$) }
        \else
        \includegraphics[width=\textwidth, rviewport= {\llx} {\lly} {\urx} {\ury}, clip]{cameraman_N65536_M25000_SNR40_adp0_rwt1_daub79}
        \captionsetup{font=footnotesize}
        \caption{PSNR: $28.17$\,dB --- ($63$) }
        \fi
        \label{fig:cameraman2}
    \end{subfigure}
    \begin{subfigure}[t]{\figwidth}
        \centering
        \ifLargeM
        \includegraphics[width=\textwidth, rviewport= {\llx} {\lly} {\urx} {\ury}, clip]{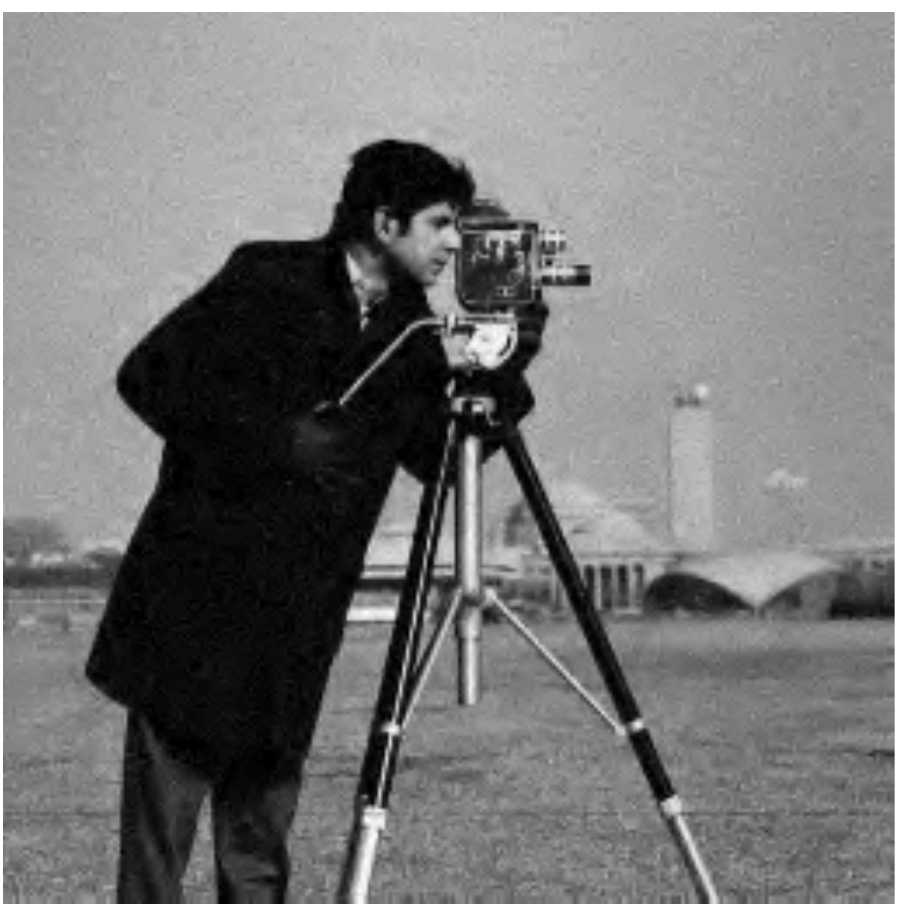}
        \captionsetup{font=footnotesize}
        \caption{PSNR: $30.27$\,dB --- ($193$) }
        \else
        \includegraphics[width=\textwidth, rviewport= {\llx} {\lly} {\urx} {\ury}, clip]{cameraman_N65536_M25000_SNR40_adp0_rwt4_daub79}
        \captionsetup{font=footnotesize}
        \caption{PSNR: $28.24$\,dB --- ($237$) }
        \fi
        \label{fig:cameraman3}
    \end{subfigure}
    \begin{subfigure}[t]{\figwidth}
        \centering
        \ifLargeM
        \includegraphics[width=\textwidth, rviewport= {\llx} {\lly} {\urx} {\ury}, clip]{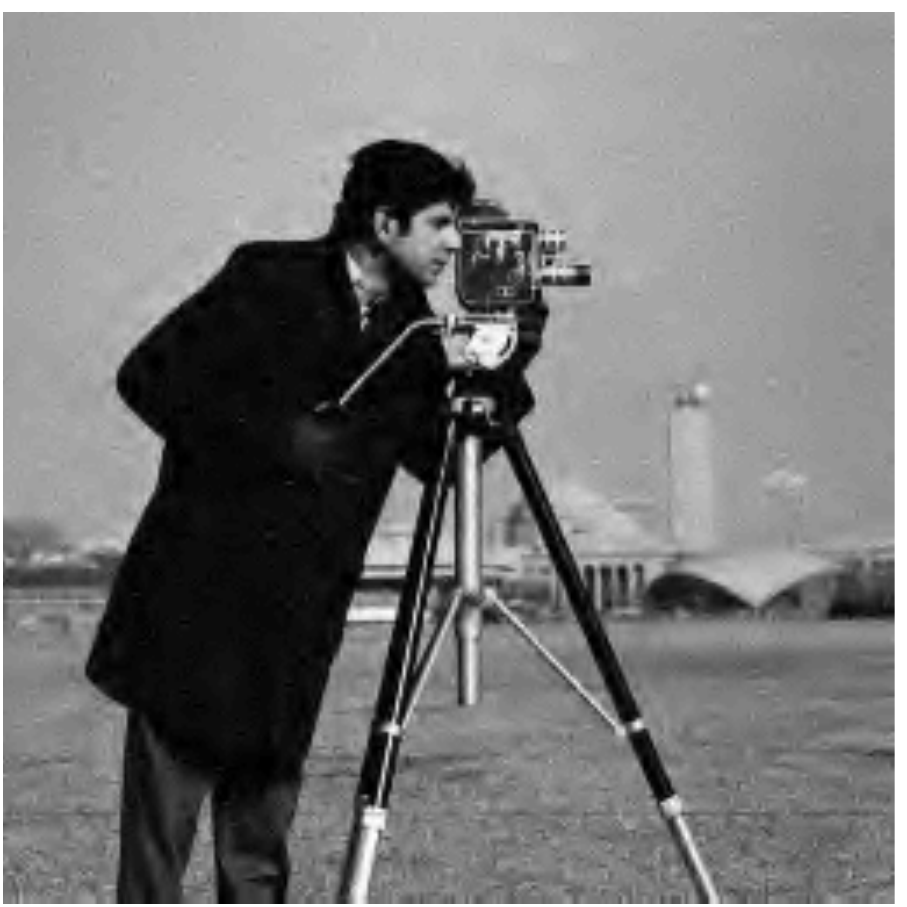}
        \captionsetup{font=footnotesize}
        \caption{PSNR: $30.77$\,dB --- ($78$)}
        \else
        \includegraphics[width=\textwidth, rviewport= {\llx} {\lly} {\urx} {\ury}, clip]{cameraman_N65536_M25000_SNR40_adp1_rwt1_daub79}
        \captionsetup{font=footnotesize}
        \caption{PSNR: $28.94$\,dB --- ($87$) }
        \fi
        \label{fig:cameraman4}
    \end{subfigure}

    ~ 
    ~ 

    \caption{Results for the recovery of $256\times256$ images from \ifLargeM $M=30,000$ \else $M = 25,000$ \fi noiselet measurements in the presence of Gaussian noise at $40${}dB SNR, using Daubechies~9/7 biorthogonal wavelet transform with odd-symmetric extensions as the sparse representation. \textbf{(Column~1)} Original images. \textbf{ (Column~2)} Portions of the images (inside the orange box) reconstructed by solving \eqref{eq:BPDN} using SpaRSA. \textbf{(Column~3)} Reconstruction after three reweighting iterations. \textbf{(Column~4)} Adaptive reweighting by updating the $\vw_i$ after every continuation step in SpaRSA. The caption under each subimage shows the PSNR over the entire reconstructed image and a count for the number of applications of $\mA^T\mA$ (in parentheses) averaged over 10 experiments.}
    \label{fig:largescale}
\end{figure*}

\fi

\bibliographystyle{IEEEtran}
\bibliography{rwtL1}

\newcommand{\noopsort}[1]{} \newcommand{\printfirst}[2]{#1}
  \newcommand{\singleletter}[1]{#1} \newcommand{\switchargs}[2]{#2#1}
\begin{thebibliography}{10}
\providecommand{\url}[1]{#1}
\csname url@samestyle\endcsname
\providecommand{\newblock}{\relax}
\providecommand{\bibinfo}[2]{#2}
\providecommand{\BIBentrySTDinterwordspacing}{\spaceskip=0pt\relax}
\providecommand{\BIBentryALTinterwordstretchfactor}{4}
\providecommand{\BIBentryALTinterwordspacing}{\spaceskip=\fontdimen2\font plus
\BIBentryALTinterwordstretchfactor\fontdimen3\font minus
  \fontdimen4\font\relax}
\providecommand{\BIBforeignlanguage}[2]{{%
\expandafter\ifx\csname l@#1\endcsname\relax
\typeout{** WARNING: IEEEtran.bst: No hyphenation pattern has been}%
\typeout{** loaded for the language `#1'. Using the pattern for}%
\typeout{** the default language instead.}%
\else
\language=\csname l@#1\endcsname
\fi
#2}}
\providecommand{\BIBdecl}{\relax}
\BIBdecl

\bibitem{Candes_2006_CompressiveSampling}
E.~Cand\`es, ``{Compressive sampling},'' \emph{Proceedings of the International
  Congress of Mathematicians, Madrid, Spain}, vol.~3, pp. 1433--1452, 2006.

\bibitem{Donoho_2006_CS}
D.~Donoho, ``Compressed sensing,'' \emph{IEEE Transactions on Information
  Theory}, vol.~52, no.~4, pp. 1289--1306, April 2006.

\bibitem{Candes_2006_StableRecovery}
E.~Cand\`es, J.~Romberg, and T.~Tao, ``{Stable signal recovery from incomplete
  and inaccurate measurements},'' \emph{Communications on Pure and Applied
  Mathematics}, vol.~59, no.~8, pp. 1207--1223, 2006.

\bibitem{CandesTao_2005_DecodingLP}
E.~Cand\`es and T.~Tao, ``Decoding by linear programming,'' \emph{IEEE
  Transactions on Information Theory}, vol.~51, no.~12, pp. 4203--4215, Dec.
  2005.

\bibitem{Tibshirani_1996_LASSO}
R.~Tibshirani, ``{Regression shrinkage and selection via the lasso},''
  \emph{Journal of the Royal Statistical Society, Series B}, vol.~58, no.~1,
  pp. 267--288, 1996.

\bibitem{Chen_99_BasisPursuit}
S.~S. Chen, D.~L. Donoho, and M.~A. Saunders, ``Atomic decomposition by basis
  pursuit,'' \emph{SIAM Journal on Scientific Computing}, vol.~20, no.~1, pp.
  33--61, 1999.

\bibitem{MeinshausenYu_2007_LassoType}
N.~Meinshausen and B.~Yu, ``{Lasso-type recovery of sparse representations for
  high-dimensional data},'' \emph{Annals of Statistics}, vol.~37, no.~1, pp.
  246--270, 2008.

\bibitem{CandesPlan_2008_NearIdealModelSelection}
E.~Cand{\`e}s and Y.~Plan, ``{Near-ideal model selection by $\ell_1$
  minimization},'' \emph{The Annals of Statistics}, vol.~37, no.~5A, pp.
  2145--2177, 2009.

\bibitem{DonohoElad_2006_StableOvercomplete}
D.~Donoho, M.~Elad, and V.~Temlyakov, ``{Stable recovery of sparse overcomplete
  representations in the presence of noise},'' \emph{IEEE Transactions on
  Information Theory}, vol.~52, no.~1, pp. 6--18, Jan. 2006.

\bibitem{zou2006adaptive}
H.~Zou, ``The adaptive lasso and its oracle properties,'' \emph{Journal of the
  American Statistical Association}, vol. 101, no. 476, pp. 1418--1429, 2006.

\bibitem{candes_enhancing_2008}
E.~J. Cand\`{e}s, M.~B. Wakin, and S.~P. Boyd, ``Enhancing sparsity by
  reweighted $\ell_1$ minimization,'' \emph{Journal of Fourier Analysis and
  Applications}, vol.~14, no. 5-6, pp. 877--905, 2008.

\bibitem{khajehnejad2010improved}
M.~Khajehnejad, W.~Xu, A.~Avestimehr, and B.~Hassibi, ``Improved sparse
  recovery thresholds with two-step reweighted $\ell_1$ minimization,'' in
  \emph{IEEE International Symposium on Information Theory Proceedings (ISIT)},
  2010, pp. 1603--1607.

\bibitem{chen2011penalized}
W.~Chen, M.~Rodrigues, and I.~Wassell, ``Penalized $\ell_1$ minimization for
  reconstruction of time-varying sparse signals,'' in \emph{Proc. {IEEE}
  International Conference on Acoustics, Speech and Signal Processing
  (ICASSP)}, 2011, pp. 3988--3991.

\bibitem{CHA-2011-SPARS}
A.~Charles and C.~Rozell, ``A hierarchical re-weighted-$\ell_1$ approach for
  dynamic sparse signal estimation,'' \emph{SPARS}, 2011.

\bibitem{FriedlanderMansour-2012-PartialSupp}
M.~Friedlander, H.~Mansour, R.~Saab, and O.~Yilmaz, ``Recovering compressively
  sampled signals using partial support information,'' \emph{IEEE Transactions
  on Information Theory}, vol.~58, no.~2, pp. 1122 --1134, Feb. 2012.

\bibitem{BergFriedlander-2008-spgl1}
E.~van~den Berg and M.~P. Friedlander, ``Probing the pareto frontier for basis
  pursuit solutions,'' \emph{SIAM Journal on Scientific Computing}, vol.~31,
  no.~2, pp. 890--912, 2008.

\bibitem{Wright-2009-sparsa}
S.~Wright, R.~Nowak, and M.~Figueiredo, ``Sparse reconstruction by separable
  approximation,'' \emph{IEEE Transactions on Signal Processing}, vol.~57,
  no.~7, pp. 2479--2493, July 2009.

\bibitem{beck-2009-FISTA}
A.~Beck and M.~Teboulle, ``A fast iterative shrinkage-thresholding algorithm
  for linear inverse problems,'' \emph{SIAM Journal on Imaging Sciences},
  vol.~2, no.~1, pp. 183--202, 2009.

\bibitem{Becker_2011_NESTA}
S.~Becker, J.~Bobin, and E.~Cand\`es., ``{NESTA: A fast and accurate
  first-order method for sparse recovery},'' \emph{SIAM Journal on Imaging
  Sciences}, vol.~4, no.~1, pp. 1--39, 2011.

\bibitem{yang-2011-yall1}
J.~Yang and Y.~Zhang, ``Alternating direction algorithms for $\ell_1$-problems
  in compressive sensing,'' \emph{SIAM Journal on Scientific Computing},
  vol.~33, no. 1-2, pp. 250--278, 2011.

\bibitem{Becker_2011_TFOCS}
S.~Becker, E.~Cand\`es, and M.~Grant, ``Templates for convex cone problems with
  applications to sparse signal recovery,'' \emph{Mathematical Programming
  Computation}, vol.~3, no.~3, 2011.

\bibitem{OsbornePresnell_2000_NewApproachLasso}
M.~Osborne, B.~Presnell, and B.~Turlach, ``{A new approach to variable
  selection in least squares problems},'' \emph{IMA Journal of Numerical
  Analysis}, vol.~20, no.~3, pp. 389--403, 2000.

\bibitem{Efron_2004_LARS}
B.~Efron, T.~Hastie, I.~Johnstone, and R.~Tibshirani, ``{Least angle
  regression},'' \emph{Annals of Statistics}, vol.~32, no.~2, pp. 407--499,
  2004.

\bibitem{AR_L1updating_JSTSP09}
M.~Asif and J.~Romberg, ``Dynamic updating for $\ell_1$ minimization,''
  \emph{{IEEE} Journal of Selected Topics in Signal Processing}, vol.~4, no.~2,
  pp. 421--434, Apr. 2010.

\bibitem{Radchenko-2011-FLASH}
P.~Radchenko and G.~James, ``Improved variable selection with forward-lasso
  adaptive shrinkage,'' \emph{The Annals of Applied Statistics}, vol.~5, no.~1,
  pp. 427--448, 2011.

\bibitem{Fuchs_2004_OnSparseRep}
J.~Fuchs, ``{On sparse representations in arbitrary redundant bases},''
  \emph{IEEE Transactions on Information Theory}, vol.~50, no.~6, pp.
  1341--1344, June 2004.

\bibitem{Donoho_2006_FastLl1}
D.~L. Donoho and Y.~Tsaig, ``Fast solution of $\ell_1$-norm minimization
  problems when the solution may be sparse,'' \emph{IEEE Transactions on
  Information Theory}, vol.~54, no.~11, pp. 4789--4812, 2008.

\bibitem{Boyd_book_ConvexOptimization}
S.~Boyd and L.~Vandenberghe, \emph{{Convex Optimization}}.\hskip 1em plus 0.5em
  minus 0.4em\relax {Cambridge University Press}, March 2004.

\bibitem{Golub_1996_MatrixComputation}
G.~Golub and C.~Van~Loan, \emph{{Matrix Computations}}.\hskip 1em plus 0.5em
  minus 0.4em\relax Johns Hopkins University Press, 1996.

\bibitem{Bjorck_1996_NumericalLS_book}
{\AA}.~Bj{\"o}rck, \emph{{Numerical Methods for Least Squares Problems}}.\hskip
  1em plus 0.5em minus 0.4em\relax Society for Industrial Mathematics, 1996.

\bibitem{afonso-2010-salsa}
M.~Afonso, J.~Bioucas-Dias, and M.~Figueiredo, ``Fast image recovery using
  variable splitting and constrained optimization,'' \emph{IEEE Transactions on
  Image Processing}, vol.~19, no.~9, pp. 2345--2356, Sept. 2010.

\bibitem{Figueiredo_2007_GPSR}
M.~Figueiredo, R.~Nowak, and S.~Wright, ``Gradient projection for sparse
  reconstruction: Application to compressed sensing and other inverse
  problems,'' \emph{IEEE Journal of Selected Topics in Signal Processing},
  vol.~1, no.~4, pp. 586--597, Dec. 2007.

\bibitem{hale2008fpc}
E.~Hale, W.~Yin, and Y.~Zhang, ``{Fixed-Point Continuation for
  $\ell_1$-minimization: Methodology and Convergence},'' \emph{SIAM Journal on
  Optimization}, vol.~19, pp. 1107--1130, 2008.

\bibitem{Wavelab}
J.~Buckheit, S.~Chen, D.~Donoho, and I.~Johnstone, ``{Wavelab 850, Software
  toolbox},'' {http://www-stat.stanford.edu/$\sim$wavelab/}.

\bibitem{AR_l1_homotopy_webpage}
M.~S. Asif and J.~Romberg, ``{$\ell_1$ Homotopy : A MATLAB toolbox for homotopy
  algorithms in $\ell_1$ norm minimization problems},''
  http://users.ece.gatech.edu/$\sim$sasif/homotopy.

\bibitem{Daubechies_2004_ItrativeThresholding}
I.~Daubechies, M.~Defrise, and C.~De~Mol, ``{An iterative thresholding
  algorithm for linear inverse problems with a sparsity constraint},''
  \emph{Communications on Pure and Applied Mathematics}, vol.~57, no.~11, pp.
  1413--1457, 2004.

\bibitem{bioucas-2007-TwIST}
J.~Bioucas-Dias and M.~Figueiredo, ``A new {TwIST}: two-step iterative
  shrinkage/thresholding algorithms for image restoration,'' \emph{IEEE
  Transactions on Image Processing}, vol.~16, no.~12, pp. 2992--3004, Dec.
  2007.

\bibitem{Mansour-2012-WSPGL1}
H.~Mansour, ``Beyond $\ell_1$-norm minimization for sparse signal recovery,''
  in \emph{Proc. of the IEEE Statistical Signal Processing Workshop (SSP)},
  August 2012.

\bibitem{CDF-1992-Biorthogonal}
A.~Cohen, I.~Daubechies, and J.-C. Feauveau, ``Biorthogonal bases of compactly
  supported wavelets,'' \emph{Communications on Pure and Applied Mathematics},
  vol.~45, no.~5, pp. 485--560, 1992.

\bibitem{Mallat_book_WaveletTour99}
S.~Mallat, \emph{A Wavelet Tour of Signal Processing, Second Edition (Wavelet
  Analysis \& Its Applications)}.\hskip 1em plus 0.5em minus 0.4em\relax
  {Academic Press}, September 1999.

\bibitem{Coifman_2001_Noiselets}
R.~Coifman, F.~Geshwind, and Y.~Meyer, ``{Noiselets},'' \emph{Applied and
  Computational Harmonic Analysis}, vol.~10, no.~1, pp. 27--44, 2001.

\end{thebibliography}
\end{document}